\documentclass[sigplan,nonacm]{acmart}
\usepackage[T1]{fontenc}
\usepackage[english]{babel}
\usepackage{graphicx}
\usepackage{url}
\usepackage{multirow}
\usepackage{amsmath}
\usepackage{amsthm}
\usepackage{xcolor}
\usepackage{xspace}
\usepackage{algorithm}
\usepackage{algpseudocode}
\usepackage{subcaption}
\usepackage{float}
\usepackage{stfloats}
\usepackage[font=small,labelfont=bf]{caption}

\usepackage{amssymb}
\usepackage{pifont}
\usepackage{enumitem}
\usepackage{makecell}
\usepackage[normalem]{ulem}
\usepackage{booktabs}
\usepackage{tabularx}
\usepackage{balance}
\usepackage{threeparttable}
\usepackage{todonotes}
\microtypecontext{spacing=nonfrench}
\setitemize{noitemsep,topsep=3pt,parsep=0em}
\DisableLigatures{encoding=*, family=*}
\newcommand{\sysname}{DecoupleVS\xspace}

\definecolor{OliveGreen}{cmyk}{0.64,0,0.95,0.40}
\definecolor{ao}{rgb}{0.0, 0.5, 0.0}
\definecolor{asparagus}{rgb}{0.53, 0.66, 0.42}
\definecolor{applegreen}{rgb}{0.55, 0.71, 0.0}
\definecolor{aogreen}{rgb}{0.0, 0.5, 0.0}
\definecolor{columbiablue}{rgb}{0.61, 0.87, 1.0}
\definecolor{cornellred}{rgb}{0.7, 0.11, 0.11}
\definecolor{cornflowerblue}{rgb}{0.39, 0.58, 0.93}
\definecolor{denim}{rgb}{0.08, 0.38, 0.74}

\colorlet{BtfulGreen}{black!30!green!70!}
\colorlet{BtfulOrange}{white!10!orange!90!}
\colorlet{BtfulGray}{white!50!gray!50!}

\newcommand{\para}[1]{\noindent\textbf{#1}}
\newcommand{\cut}[1]{}

\renewcommand\footnotetextcopyrightpermission[1]{}

\begin{document}

\title{Decoupling Vector Data and Index Storage for Space Efficiency}
\author{Yuanming Ren$^1$, Juncheng Zhang$^1$, Yanjing Ren$^1$, Rui Yang$^2$, Di Wu$^2$, and
Patrick P. C.  Lee$^1$
\vspace{3pt}\\
\emph{$^1$The Chinese University of Hong Kong} \ \ 
\emph{$^2$Bytedance}}

\maketitle

\begin{sloppypar}
\section*{Abstract}
Managing large-scale vector datasets with disk-resident graph approximate nearest neighbor search (ANNS) systems incurs substantial storage overhead due to the co-location of vector data and auxiliary index metadata, which prevents the storage layer from exploiting their distinct compressibility. We present \sysname, a component-aware compressed storage framework for disk-resident graph vector search. Leveraging data-index decoupling as a foundation, \sysname losslessly compresses each component according to its distinct compressibility characteristics, thereby significantly reducing storage space. It further adapts the search and update paths to preserve their performance under compressed storage layouts.  Evaluation on real-world public and proprietary billion-scale datasets shows that \sysname reduces storage space by up to 58.7\%, while delivering improved or competitive search and update performance compared to state-of-the-art disk-resident graph ANNS systems.


\section{Introduction}
\label{sec:intro}

The rapid growth of AI applications, including web search \cite{grbovic18}, recommendation \cite{covington16}, and retrieval-augmented generation \cite{chatgpt}, has spurred demand for managing increasingly large collections of high-dimensional {\em vectors}. These vectors encode unstructured data (e.g., text, images, audio, video) in a common embedding space, making similarity search a core primitive in modern data-intensive systems. For example, graph-based approximate nearest neighbor search (ANNS) organizes vectors as vertices and connects similar vectors with edges, and improves search efficiency by returning an approximate top-$K$ result set instead of exact neighbors, trading a small accuracy loss for large performance gains. 

A vector dataset typically comprises both {\em vector data} (the data component), full-precision vectors that store embedding values for accurate distance computation, and an {\em auxiliary index} (the metadata component), which stores per-vector metadata for efficient ANNS. As vector datasets scale to hundreds of trillions of vectors \cite{trillion-scale}, in-memory ANNS systems (e.g., HNSW \cite{malkov18}) become impractical due to prohibitive memory requirements \cite{zhang24}. This motivates {\em disk-resident} graph ANNS systems (e.g., \cite{jayaram19, chen21, xu23, tian25}) that offload vector datasets to solid-state drives (SSDs) or other persistent storage media, thereby enabling billion-scale deployments. 


Existing disk-resident graph ANNS systems are mainly optimized for search efficiency by co-locating vector data and auxiliary index metadata in storage. This simplifies implementation and amortizes I/Os across vector data and metadata access in a single read.  However, this search-friendly layout is {\em storage-inefficient}. First, the total size of graph metadata and vector data can be larger than the raw data (\S\ref{subsec:limitations}). Second, fixed-size page-aligned records can suffer from internal fragmentation when the combined vector data and metadata size does not align with the storage page size. Third, co-locating vector data and metadata forces the storage layer to treat semantically distinct components as a single opaque record, preventing tailored compression for each component.



In this paper, we study the storage-compression problem in disk-resident graph ANNS systems. Our insight is that vector data and auxiliary index metadata exhibit fundamentally distinct characteristics in data semantics and access patterns (\S\ref{subsec:motivation}), enabling {\em component-aware compression} to achieve substantially higher compression ratios than general-purpose alternatives.  A natural way to exploit these differences is to separate vector data from auxiliary index metadata, allowing each to be compressed, laid out, and accessed independently. While data-index decoupling has also been considered concurrently by prior work \cite{lou26,zhao26} (\S\ref{sec:related}), applying this idea to storage compression introduces three non-trivial challenges.  First, general-purpose lossless compression techniques (e.g., LZ4 \cite{lz4}, Huffman \cite{knuth85}) should be adopted for vector storage to preserve data fidelity, yet they fail to exploit the semantic structure of vector data and auxiliary index metadata; improper use of them yields suboptimal compression gains. Second, compression produces variable-size data that cannot be directly addressed by vector ID, requiring additional metadata management to support efficient random access. Third, separating vector data and auxiliary index metadata into independent physical storage introduces I/O locality loss during search and consistency challenges during updates.

We design \sysname, a \uline{decouple}d \underline{v}ector \underline{s}torage management framework for disk-resident graph ANNS systems. Using data-index decoupling as an enabling substrate, \sysname adopts a co-design of lossless compression, data layouts, and search and update paths without compromising performance. Instead of proposing new compression primitives, \sysname shows how compression can be made practical for compressed decoupled storage and random-access graph ANNS workloads through: (i) component-aware lossless encoding of vector data and auxiliary index metadata, (ii) hierarchical layouts that enable direct access over variable-size compressed records, (iii) latency-aware search that removes vector data reads from the traversal critical path, and (iv) decoupled update paths that combine graph batch merges with log-structured vector data storage.

We conduct experiments on real-world public and proprietary datasets.  Compared to state-of-the-art disk-resident graph ANNS systems (DiskANN \cite{jayaram19} and PipeANN \cite{guo25}), \sysname saves up to 58.7\% of storage space, achieves up to 2.39$\times$ search throughput gain, and preserves high update efficiency. We will open-source \sysname in the final paper. 

\section{Background}
\label{sec:background}


\subsection{Basics of ANNS}
\label{subsec:basics}

An ANNS system identifies the most similar items in a large vector dataset for a search query.  It employs deep-learning-based embedding models to map unstructured data into high-dimensional vectors within a common vector space, and constructs an auxiliary index over these vectors for efficient similarity search.  Given a search query, the system generates a query vector and uses the auxiliary index to locate the most similar vectors according to a distance metric (e.g., Euclidean distance, inner product, or cosine similarity), then returns the top-$K$ items corresponding to the nearest vectors. Each vector typically comprises hundreds of dimensions encoded as numerical data types, such as \texttt{UINT8}, \texttt{INT8}, or \texttt{FP32} \cite{annbenchmarks}, where each dimension captures a specific semantic feature learned by the embedding model \cite{reimers19}. To maintain search accuracy, vectors are often normalized to prevent large-value dimensions from dominating distance calculations.  Vector dimensionality can reach tens of thousands when representing rich textual data (e.g., in large language models) \cite{brown20}.

The auxiliary index enables efficient search without scanning the entire dataset, and can be structured as graph-based \cite{malkov18, jayaram19}, cluster-based \cite{jegou10,chen21}, or tree-based \cite{flann}.  For example, a graph-based index represents each vector as a vertex and connects similar vertices by edges.  Given a search query, the system begins at a designated entry point, traverses the graph along edges, and computes distances between the query vector and visited vertices to construct a candidate set. If no closer vectors can be found, it returns the top-$K$ candidates as the {\em result set}. In-memory ANNS systems (e.g., HNSW \cite{malkov18}, IVFADC \cite{jegou10}, and RUMMY \cite{zhang24}) store both vectors and auxiliary indexes in DRAM for fast access, but incur substantial memory overhead (e.g., RUMMY \cite{zhang24} requires terabytes of memory to manage billion-scale datasets), motivating the adoption of disk-resident ANNS systems. 

\subsection{Disk-Resident ANNS Systems}
\label{subsec:limitations}

To reduce memory footprints, disk-resident ANNS systems (e.g., \cite{jayaram19, chen21, xu23, tian25}) adopt a hybrid storage architecture that retains compact in-memory representations while persisting full-precision vector data and auxiliary index metadata on disk.  For example, DiskANN \cite{jayaram19} keeps lossy-compressed vectors in memory (via product quantization (PQ) \cite{jegou10}), while SPANN \cite{chen21} keeps an in-memory graph of cluster centroids, each linked to an on-disk posting list of full-precision vector data.  However, SPANN replicates each vector up to eight times across different posting lists to maintain high search accuracy, incurring substantial storage overhead (\S\ref{subsec:overall}). Given that DiskANN exhibits a more space-efficient design, we focus on it for problem motivation.

\begin{figure}[!t]
\centering
\includegraphics[width=0.97\linewidth]{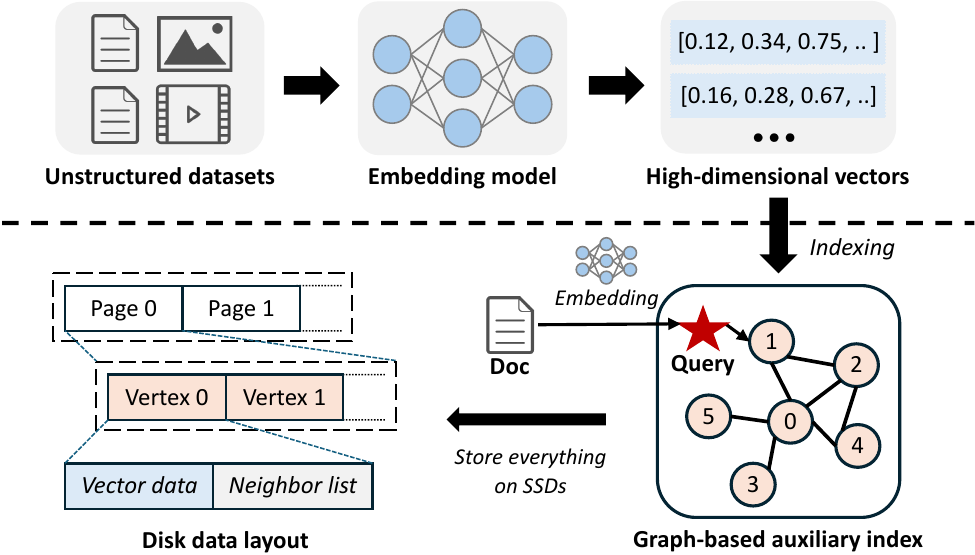}
\vspace{-9pt}
\caption{A disk-resident graph ANNS system based on DiskANN.} 
\label{fig:basics}
\vspace{-9pt}
\end{figure}

Figure~\ref{fig:basics} depicts the storage layout of a disk-resident graph ANNS system based on DiskANN, with graph index vertices sorted by ID and stored sequentially.  Each vertex bundles a full-precision vector with its neighbor list, enabling both to be fetched in a single disk I/O during graph traversal. Since vertices are fixed-size and page-aligned, DiskANN can directly compute any vertex's disk offset from its ID without additional metadata lookups. For a search query, DiskANN maintains a candidate list of size $L$, initialized with the neighbors of a designated entry point ranked by their PQ-based distances to the query vector. It then expands the search frontier iteratively: in each iteration, it issues batch I/O requests to fetch the top-$W$ unvisited candidates ($W$ is the {\em beam width}) and evaluates their neighbors via in-memory PQ codes to select the most promising candidates without reading full-precision vectors from disk.  Finally, it re-ranks the visited vertices using their full-precision vectors to return the top-$K$ result set. Despite various I/O optimizations, existing disk-resident ANNS systems still face two critical limitations at the persistent storage layer.

\para{Limitation \#1: High storage costs.} Disk-resident graph ANNS systems incur huge storage overhead for persisting both high-dimensional vector data and auxiliary index metadata.  For instance, a 128-dimensional vector \cite{sift} with each dimension encoded as a 4-byte float consumes 512~bytes.  The auxiliary index connecting each vector to 128 neighbors \cite{guo25} adds another 512~bytes of 4-byte integer neighbor IDs, bringing the total storage per vector to 1\,KiB (i.e., doubling the raw vector size).  This overhead compounds severely at scale: vector datasets can grow to hundreds of trillions of vectors in production \cite{trillion-scale}, and the footprint is further amplified by replication for fault tolerance and load balancing \cite{anns-replication}.


\begin{figure}[!t]
\centering
\includegraphics[width=0.97\linewidth]{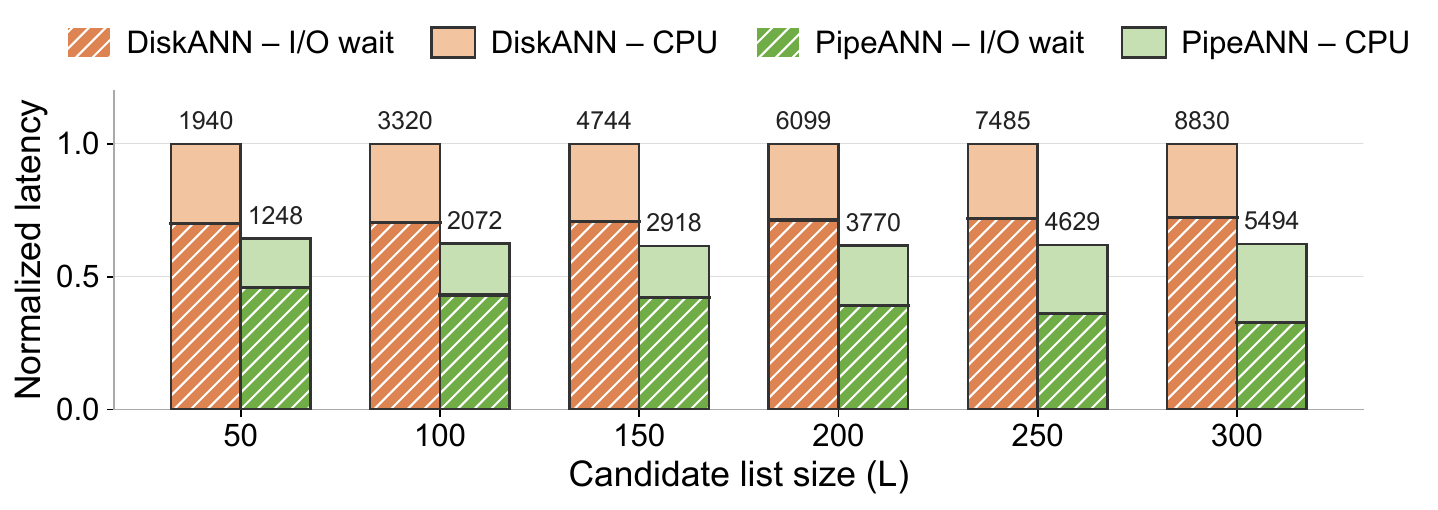}
\vspace{-9pt}
\caption{
Search latency breakdown of DiskANN and PipeANN (CPU time vs. I/O wait time), normalized to DiskANN's latency. Numbers above bars are absolute latencies (\textmu{}s).} 
\label{fig:iowait}
\vspace{-6pt}
\end{figure}

\para{Limitation \#2: High I/O demands.}  Disk-resident graph ANNS systems suffer from significant I/O amplification in both search and update operations.  Search queries must traverse large portions of the auxiliary index on disk to identify the result set. This disk-bound traversal can account for over 70\% of the total query time \cite{liang22}, and this bottleneck persists even with pipelined I/O and computation \cite{guo25}.  To verify, we profile DiskANN \cite{jayaram19} and PipeANN \cite{guo25} on the SIFT1M dataset with 1\,M 128-dimensional vectors\footnote{We choose this small-scale dataset for ease of analysis and visualization, yet our findings hold consistently for billion-scale datasets (\S\ref{sec:evaluation}).} \cite{sift}. We measure the {\em I/O wait time} (i.e., the duration that CPU threads are idle awaiting disk I/Os) with beam width $W=4$ (as recommended in \cite{jayaram19}) while varying the candidate list size $L$ from 50 to 300 to cover various search accuracy levels. Figure~\ref{fig:iowait} shows that even with pipelining, PipeANN's I/O wait time still accounts for 52.9-72.2\% of overall search latency, confirming that the I/O demands of graph traversal cannot be fully alleviated by pipelining alone.

Updates are also costly: each vector update modifies not only vector data but also all relevant auxiliary index metadata.  For example, for a graph-based index, updating a vector also revises the adjacency lists of all its neighbors.  Even with buffered updates \cite{singh21}, the auxiliary index still needs to be periodically rewritten to disk in full to incorporate accumulated changes, consuming substantial write bandwidth.   


\subsection{Motivation and Challenges}
\label{subsec:motivation}

Vector data and auxiliary index metadata have distinct characteristics. We identify two key differences that lead to storage and I/O inefficiencies under co-located storage.


{\em (i) Different data semantics.} Vector data generated by embedding models encodes high-dimensional numerical features with dimension-wise statistical regularity, while auxiliary index metadata primarily contains neighbor IDs as fixed-length integers (e.g., adjacency lists in graph-based indexes \cite{jayaram19} or posting lists in a cluster-based index \cite{chen21}).  Co-locating them in storage forfeits opportunities for component-aware optimizations on each. 

{\em (ii) Different access patterns.} During search, the auxiliary index is traversed frequently to identify candidate vectors, while vector data is accessed only sporadically for final re-ranking \cite{jayaram19} (using full-precision vectors for accurate distance computation).  During updates, auxiliary index metadata must be kept current to maintain search accuracy \cite{xu23,xu25}, while vector data can be updated asynchronously without compromising correctness.  Such distinct access patterns motivate independent I/O handling for them.

\para{Design considerations.} To address the above storage and I/O inefficiencies, we explore decoupled storage management for vector data and auxiliary index metadata.  This raises four design questions.

{\em Q1: How to maximize storage savings?} Lossless compression is necessary to preserve data integrity: full-precision vectors are required for accurate final re-ranking \cite{jayaram19}, and the auxiliary index must remain exactly correct for search algorithms. However, general-purpose lossless techniques, such as entropy coders (e.g., Huffman \cite{knuth85}, ANS \cite{duda13}) and dictionary coders (e.g., LZ77 \cite{rigler07}, LZ4 \cite{lz4}), treat data as opaque byte streams and fail to exploit the semantic structure of vectors or the integer-sequence regularity of auxiliary index metadata, yielding suboptimal compression ratios.

{\em Q2: How to mitigate the operational overhead on compressed data?} Compression introduces two forms of overhead. First, decompression at query time adds computational costs that must not degrade search or update performance. Second, compression produces variable-size outputs, eliminating the fixed-stride layout that allows direct disk addressing by vector ID. Variable-size data requires additional metadata management and indirection to support efficient random access.


{\em Q3: How to maintain high search performance and accuracy?} Separating vector data and auxiliary index metadata in physical storage eliminates I/O locality, forcing multiple I/Os to distinct storage locations per graph traversal step. This extra I/O overhead risks degrading search performance; any mitigation strategy must preserve recall accuracy to the same level as co-located designs. 


{\em Q4: How to support updates under a decoupled storage layout?} Ensuring consistency across decoupled components requires coordinated updates to both vector data and auxiliary index metadata on disk \cite{singh21, guo26}.  While decoupling breaks spatial locality, it also creates an opportunity to handle updates for each component independently, potentially reducing overall write amplification.

\section{\sysname Design}
\label{sec:design}


\subsection{Design Overview}
\label{subsec:overview}

\begin{figure}[!t]
\centering
\includegraphics[width=0.99\linewidth]{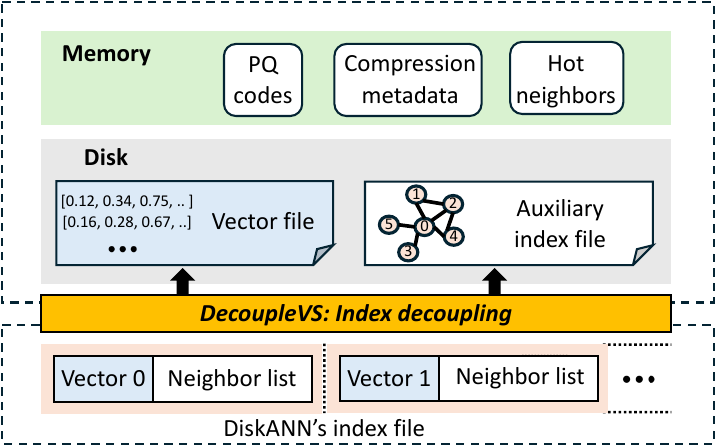}
\vspace{-6pt}
\caption{Architecture of \sysname.}
\label{fig:arch}
\vspace{-9pt}
\end{figure}

\sysname is a compressed storage framework for disk-resident graph ANNS systems. Its primary objective is to reduce the persistent storage footprint of vector data and auxiliary index metadata while preserving the random-access interface for existing graph search and update algorithms. Figure~\ref{fig:arch} shows \sysname's architecture, which builds on DiskANN \cite{jayaram19}. \sysname stores the auxiliary index (including each vertex's neighbor count and neighbor list) and vector data in separate on-disk files; this decoupled organization is not the main contribution by itself, but rather enables \sysname to apply component-aware compression. Like DiskANN, \sysname caches PQ-compressed vectors in memory; additionally, it maintains compression metadata and hot neighbor lists in memory to support efficient compressed storage access (\S\ref{subsec:search}). \sysname introduces four techniques to optimize space efficiency, search, and updates.
\begin{itemize}[leftmargin=*]
\item 
{\em Tailored lossless compression} (\S\ref{subsec:compression}).  \sysname applies component-aware lossless compression schemes to vector data and auxiliary index metadata, exploiting their distinct compressibility characteristics.
\item 
{\em Hierarchical storage layouts} (\S\ref{subsec:layout}). \sysname organizes compressed vector data and auxiliary index metadata in hierarchical on-disk layouts that enable efficient random access and updates while bounding the in-memory compression metadata footprint.
\item 
{\em Latency-aware vector search} (\S\ref{subsec:search}).  \sysname assigns different I/O paths to vector data and auxiliary index metadata based on their distinct roles in search, keeping index access on the critical path while offloading vector fetching. It also uses adaptive prefetching to avoid unnecessary I/O.
\item 
{\em Decoupled vector updates} (\S\ref{subsec:update}). \sysname separates the update paths for vector data and auxiliary index metadata. It batch-merges index updates to preserve graph integrity \cite{singh21}, while appending vector data in a log-structured manner and deferring space reclamation to background garbage collection, achieving low write amplification.
\end{itemize}

\para{Applicability.} \sysname is built and evaluated for disk-resident graph ANNS systems derived from DiskANN. It demonstrates how separating vector data from auxiliary index metadata enables component-aware compression and access optimizations, and we expect its core design principles to extend beyond the current implementation. In particular, vector data compression is independent of the graph structure and applies to systems that store full-precision vectors on disk, while auxiliary index metadata compression exploits a common property that both graph-based (neighbor lists) and cluster-based (posting lists) systems index vectors by integer identifiers that can often be reordered into monotonically increasing sequences for compact encoding.  Nevertheless, cluster-based systems (e.g., SPANN) have different access patterns and posting-list layouts. Extending \sysname's design to those systems requires additional engineering and is left for future work.

\subsection{Tailored Lossless Compression}
\label{subsec:compression}

\begin{table}[!t]
\small
\setlength{\tabcolsep}{2.5pt}
\centering
\renewcommand{\arraystretch}{1.1}
\caption{Characterization of different datasets.}
\label{tab:vec_character}
\vspace{-9pt}
\begin{tabular}{|l|c|c|c|}
\hline
\textbf{Metric} & \textbf{SIFT1M} & \textbf{SPACEV1M} & \textbf{DecoupleVS1M} \\
\hline
Global dispersion      & 36.2  & 12.2  & 0.09 \\
\hline
Dimensional dispersion & 32.8  & 9.69  & 0.06 \\
\hline
Global entropy  & 2.63  & 5.59  & 4.39 \\
\hline
Columnar entropy& 1.73  & 5.46  & 2.86 \\
\hline
\end{tabular}
\vspace{-9pt}
\end{table}


\noindent{\bf Vector data characterization.}  We characterize the compressibility of vector data and auxiliary index metadata using three 1\,M-vector datasets: (i) SIFT1M \cite{sift} with 128 dimensions, (ii) SPACEV1M \cite{spacev} with 100 dimensions, and (iii) DecoupleVS1M, a 128-dimension proprietary dataset from a commercial application (billion-scale validation is deferred to \S\ref{sec:evaluation}). Since vectors are normalized (\S\ref{subsec:basics}), we hypothesize that values within the same dimension exhibit lower variation than the global distribution.  We measure this via: (i) {\em global dispersion}, the standard deviation across all values in the dataset, and (ii) {\em dimensional dispersion}, the average per-dimension standard deviation.  Table~\ref{tab:vec_character} confirms that dimensional dispersion is consistently lower than global dispersion across all three datasets, indicating that values within each dimension are more concentrated and hence more compressible. We further quantify compressibility via information entropy \cite{shannon48} based on: (i) {\em global entropy}, computed across all bytes in the dataset, and (ii) {\em columnar entropy}, the average entropy of each byte column across all vectors.  Table~\ref{tab:vec_character} shows that columnar entropy is consistently lower than global entropy across all datasets, by 34.3\% and 34.8\% for SIFT1M and DecoupleVS1M, respectively.  This indicates strong byte-positional locality that is exploitable for compression.

\para{Delta compression for multi-dimensional vectors.} Our entropy analysis shows that compressing values at the same byte position across vectors yields high savings. However, doing so directly requires a separate compression stream per byte position, causing numerous decompression operations that severely degrade retrieval performance. Instead, \sysname constructs a base vector by choosing the most frequent byte value at each byte position across all vectors being considered.  It then applies XOR-based delta compression \cite{hershcovitch25, wang25}: each vector is bitwise-XORed with the base vector to produce a low-entropy XOR-delta. The XOR-deltas are then losslessly compressed using Huffman coding \cite{knuth85}, which offers fast decoding and near-optimal compression ratios \cite{entropycoders}.  This preserves byte-positional locality while enabling vector-level compression through a single, unified compression stream. Compared to prior delta compression approaches \cite{ning24, hershcovitch25, wang25}, which operate on one-dimensional byte sequences, \sysname explicitly handles multi-dimensional vectors by constructing a dimension-aligned base vector. 


\para{Auxiliary index compression.} Each vector's auxiliary index metadata comprises a neighbor count and an adjacency list of integer IDs.  Since all neighbors are evaluated during search in an order-independent manner, \sysname sorts neighbor IDs in ascending order and transforms each adjacency list into a monotonically increasing integer sequence, making it highly compressible.  It then applies the {\em Elias-Fano} algorithm \cite{elias74,fano71} to encode the sorted list using a {\em two-level representation}: the lower bits of each ID are stored at full precision with a fixed width across all IDs, while the higher bits of all IDs are encoded collectively in a compact bitmap.

\subsection{Hierarchical Storage Layouts}
\label{subsec:layout}

\sysname introduces hierarchical storage layouts for vector data and auxiliary index metadata to enable efficient random access and updates to variable-size compressed data.

\begin{figure}[!t]
\centering
\includegraphics[width=0.99\linewidth]{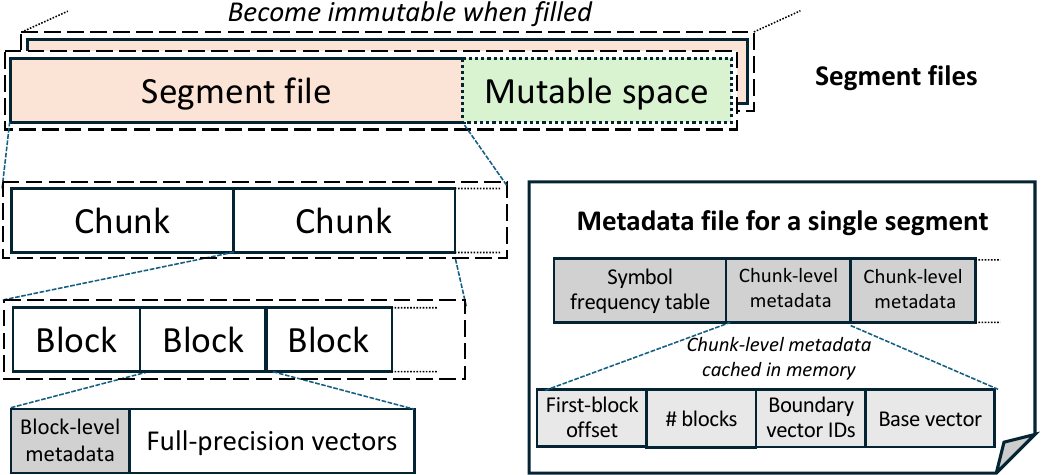}
\vspace{-6pt}
\caption{Hierarchical storage layout for vector data.}
\label{fig:vector_storage}
\vspace{-9pt}
\end{figure}

\para{Segment-level vector data organization.} As shown in Figure~\ref{fig:vector_storage}, \sysname organizes vector data into fixed-capacity {\em segment} files to bound the maximum number of vectors being stored. A segment is initially mutable and accepts new vectors via log-structured writes.  Once a segment reaches capacity, it is sealed and made immutable; a background thread pool then compresses it in parallel.  Segment IDs start from 0, and a given vector ID is mapped to its segment by integer division of the vector ID by the segment capacity.  During updates, \sysname asynchronously frees invalid vectors once the stale data volume in a segment exceeds a predefined threshold (\S\ref{subsec:update}), reclaiming space without rewriting the entire index as in prior work \cite{singh21}. 


Each compressed segment is subdivided into fixed-capacity {\em chunks}.  In each chunk, vectors are packed into 4\,KiB physical disk blocks, the minimum I/O units, and sorted by IDs.  To locate the block containing a given vector ID, \sysname maintains in-memory chunk-level metadata, including the offset of the chunk's first block (4~bytes), the number of blocks (4~bytes), the boundary vector IDs of all blocks (4~bytes each), and a base vector for delta compression (\S\ref{subsec:compression}). All chunk-level metadata for a segment is persisted to a separate disk file that is loaded into memory at startup for fast lookups without scanning the segment. To retrieve a vector, \sysname queries the in-memory chunk metadata to locate the target block, issues a single disk read for the block, and decompresses the target vector. This three-level hierarchy (segment$\rightarrow$chunk$\rightarrow$block) bounds the metadata memory footprint while preserving efficient random access.

\para{Segment-level vector compression.} \sysname's Huffman-based compression (\S\ref{subsec:compression}) requires a symbol frequency table. A single global table over the entire dataset ignores local data statistics, while per-chunk tables incur excessive metadata overhead. Thus, \sysname builds one table per segment, balancing compression fidelity and metadata costs.  It performs compression in two stages. In the first stage, it determines whether to apply XOR-based delta compression (\S\ref{subsec:compression}) to each chunk.  It samples a portion of the vectors (e.g., the first 10\%) and computes the entropy of both the raw vectors and their XOR-deltas against a candidate base vector, built by selecting the most frequent byte value at each byte position within the chunk, and applies delta compression to the whole chunk only if it yields lower entropy. This ensures that delta compression is applied only when it achieves storage savings. \sysname stores the outputs, along with the base vector (if used), in the chunk-level metadata.  In the second stage, \sysname constructs a unified symbol frequency table from all chunks in the segment and encodes every vector using the resulting Huffman code.  The frequency table is persisted alongside the chunk-level metadata (Figure~\ref{fig:vector_storage}), and is reconstructed in memory at segment load time and cached for fast decompression during search. 

\para{Block-based auxiliary index organization.} 
\sysname stores the compressed auxiliary index in a block-based layout analogous to that of vector data: each 4\,KiB disk block holds multiple compressed adjacency lists preceded by a block-level metadata header. To enable fast random access, \sysname maintains a sparse in-memory index mapping boundary vector IDs to block offsets, so that any target adjacency list can be located with a single lookup. This sparse index is persisted to disk and loaded into memory at startup.

\para{Compression metadata memory management.}  \sysname keeps three types of compression metadata in memory: (i) chunk-level metadata, (ii) per-segment frequency tables, and (iii) the sparse auxiliary index. The chunk-level metadata footprint depends on chunk size $C$, creating a trade-off between memory usage and access granularity. \sysname exposes a configurable parameter $\beta$, the chunk-metadata overhead ratio (chunk metadata size / total vector data size). Let $V$ be the full-precision vector size and $\alpha\in(0,1]$ be the compression ratio (compressed / uncompressed size; smaller $\alpha$ means greater savings). The expected number of 4\,KiB blocks per chunk is $\tfrac{\alpha \times C}{4096}$, giving a per-chunk metadata size of $4\times \left(\tfrac{\alpha \times C}{4096} + 3\right) + V$~bytes. This yields $\beta = \tfrac{V + 12}{C} + \tfrac{\alpha}{1024}$. Given a user-configured $\beta$ and known $\alpha$ and $V$, \sysname solves for the required $C$. When $\alpha$ is unknown, setting $\alpha = 1$ gives a conservative upper bound (i.e., no compression savings). 

The sparse in-memory index size is determined by the number of blocks in the compressed auxiliary index.
Specifically, for a neighbor list of degree $R$ and a maximum vector ID $N$, the worst-case encoded size is at most $2R + R\cdot\lceil \log_2 \tfrac{N}{R} \rceil$~bits \cite{ottaviano14}. With $N$ such lists across 4\,KiB blocks, the worst-case block count is $\lceil \frac{N(2R + R\lceil \log_2 (N/R) \rceil)}{4096 \times 8} \rceil$. The sparse in-memory index uses 4~bytes per entry, totaling $\lceil \frac{N(2R + R\lceil \log_2 (N/R) \rceil)}{8192} \rceil$~bytes. For instance, with $R=96$ and $N=10^8$, the worst-case sparse in-memory index size is about 24.6\,MiB. This is consistent with the 19.6\,MiB measured on SIFT100M (\S\ref{subsec:update}), confirming that practical encoding stays well below the bound. Additionally, the per-segment frequency tables have fixed, workload-independent sizes that are small in practice (\S\ref{sec:evaluation}).

\subsection{Latency-aware Vector Search}
\label{subsec:search}

Since the decoupled organization causes search operations to issue separate I/Os to vector data and auxiliary index metadata, \sysname employs two complementary strategies to preserve graph traversal performance over compressed decoupled storage: (i) {\em differentiated I/O paths} that keep latency-critical index access on the fast path, and (ii) {\em adaptive vector prefetching} that avoids unnecessary vector I/Os.

\para{Differentiated I/O paths.} \sysname pipelines I/Os and computations (including decompression and distance calculations), as in PipeANN \cite{guo25}, to maximize CPU and I/O utilizations.  However, as shown in \S\ref{subsec:limitations}, I/O wait time still dominates even with pipelining. The root asymmetry is that graph traversal throughput is limited by neighbor-list access latency, whereas vector data access affects only the recall accuracy of the final re-ranking.  Thus, \sysname removes vector data I/Os from the critical search path entirely: during graph traversal, it reads only the compressed auxiliary index to identify candidate vector IDs, caching hot compressed neighbor lists in memory under the least-recently-used (LRU) policy; it separately prefetches the necessary vector data only for final re-ranking.

Since compressed neighbor lists are variable-size, \sysname allocates fixed-size LRU cache entries sized to the Elias-Fano worst-case bound (\S\ref{subsec:layout}): $2R + R\cdot\lceil \log_2 \tfrac{N}{R} \rceil$ bits per list \cite{ottaviano14}, versus $32(R+1)$ bits for an uncompressed list (a vertex and its $R$ neighbors). At $R=128$ and $N=10^9$ (\S\ref{subsec:method}), this gives 2,430~bits versus 3,072~bits, achieving at least 20.9\% space reduction. By using fixed-size entries, \sysname reserves sufficient space for any compressed neighbor list, while eliminating variable-size allocation overhead.


\para{Adaptive vector prefetching.} \sysname prefetches the most promising full-precision vectors during graph traversal and terminates re-ranking early once the result set stabilizes:
\begin{itemize}[leftmargin=*]
\item 
{\em Phase 1 (Vector prefetching):} \sysname maintains a max-heap of size $K+B$ to track the top candidates by PQ-encoded distance during traversal, where $K$ is the result set size and $B$ is the re-ranking batch size (also used as the prefetch stability threshold).  Once the heap is full and $B$ consecutive candidates are explored without displacing any heap entry, the search is considered stable; using $B$ for both the stability threshold and the re-ranking batch size ensures the prefetch trigger aligns with the re-ranking granularity. \sysname then issues prefetch I/Os for the top $K$ candidates using spare I/O bandwidth, computed as the beam width $W$ \cite{jayaram19} minus the number of in-flight traversal I/Os, to avoid contention with ongoing traversal.  
\item
{\em Phase 2 (Adaptive re-ranking termination):} \sysname immediately begins re-ranking the prefetched vectors 
while concurrently issuing I/Os for the next batch. For each batch, it computes a {\em benefit ratio}, the fraction of candidates that displace an entry in the final top-$K$ heap. It terminates re-ranking if the ratio falls below a predefined threshold (default: 0.01), ensuring that I/Os are issued only while the result set continues to improve. 
\end{itemize}

\subsection{Decoupled Vector Updates}
\label{subsec:update}

\sysname preserves storage savings when vectors are inserted or deleted under compressed decoupled storage. It handles vector insertions and deletions through a decoupled update strategy that exploits the asymmetric update requirements of the two storage components: auxiliary index metadata encodes a highly interconnected graph structure that requires periodic global repair to preserve search accuracy; in contrast, vector data has no inter-vector dependencies as each vector is accessed by ID and can be written independently, making it a natural fit for log-structured storage. Specifically, \sysname applies {\em batch merges} to the auxiliary index as in FreshDiskANN \cite{singh21}, while writing vector data via segment-level appends (\S\ref{subsec:layout}). It defers space reclamation to background garbage collection (GC) to eliminate full index rewrites at each merge cycle \cite{singh21}.  Note that the auxiliary index path can be adapted to in-place techniques used by IP-DiskANN \cite{xu25} and OdinANN \cite{guo26}, as they all operate on the same underlying graph structure, while the vector data path is independent of the auxiliary index and naturally benefits from log-structured writes.

\para{Vector insertions.} \sysname inserts new vectors into the in-memory Vamana index \cite{jayaram19}. When the in-memory index reaches a capacity threshold, it computes neighbor deltas for all valid vectors and applies them to the on-disk auxiliary index in the background. Concurrently, it appends new vectors to the tail of the active mutable segment, and maintains an ID-to-location mapping within each segment group for subsequent lookups and GC. 

\para{Vector deletions.} For the auxiliary index, \sysname batches deletions and removes references to deleted vectors from their neighbors' adjacency lists to preserve graph connectivity.  For vector data, \sysname marks deleted vectors stale and reclaims their space asynchronously by GC.  GC is triggered in the background when buffered updates are flushed to disk, to limit interference with foreground operations.  It selects segments greedily by {\em garbage ratio} (i.e., fraction of stale vectors) to maximize reclaimed space per I/O. For each selected segment, \sysname copies all valid vectors into a new segment, atomically updates in-memory metadata to redirect lookups to the new locations. It also compresses and seals filled segments during compaction inline (\S\ref{subsec:layout}). The original stale segments are not released until the index switch completes, since they still serve in-flight foreground queries against the old layout.

\para{Consistency model.} \sysname follows a \emph{batch-visible} update model similar to FreshDiskANN \cite{singh21}. 
When the current batch of updates is buffered in memory, search queries run against the previous on-disk index and segments. This ensures that newly deleted vectors are never returned, even before their on-disk references are removed. Once the auxiliary-index merge and the subsequent GC cycle complete in the background, \sysname waits for all in-flight queries on the old state to finalize. It then reloads the rebuilt index and the new segment layout, and resumes serving subsequent queries against the new state. The replaced stale segments are safely released only after this switch, without exposing inconsistent vector/index states to concurrent queries.




\section{Evaluation}
\label{sec:evaluation}

We implement \sysname in C++ based on the PipeANN codebase \cite{guo25}.  PipeANN extends DiskANN \cite{jayaram19} by employing \texttt{io\_uring} as the I/O engine with asynchronous I/O requests to reduce I/O wait time; its codebase comprises 73.8\,K LoC.  We extend PipeANN with 19.1\,K LoC to support vector storage decoupling, lossless compression, and efficient operations on the compressed auxiliary index.  For Huffman coding, we adopt a high-performance implementation \cite{entropycoders} that exploits CPU out-of-order execution to accelerate compression and decompression.  For Elias-Fano encoding, we build on the Succinct library \cite{Succinct}.  We parallelize compression and decompression for the auxiliary index to boost throughput.

We evaluate \sysname with the following questions, centered on whether component-aware compression can reduce storage footprint without compromising the performance of disk-resident graph ANNS systems: 
\begin{itemize}[leftmargin=*]
\item How do \sysname's individual design components contribute to its search performance? (\S\ref{subsec:micro})
\item How does \sysname perform in terms of storage savings, search performance, and update performance? Can it scale to billion-scale datasets? (\S\ref{subsec:overall})
\item Does \sysname's tailored compression provide a better storage-performance trade-off for disk-resident graph ANNS than general-purpose compression schemes? (\S\ref{subsec:exp_analysis})
\end{itemize}

\subsection{Methodology}
\label{subsec:method}

\noindent{\bf Testbed.} We conduct experiments on a physical server equipped with a 32-core 2.30\,GHz Intel Xeon Platinum 8336C CPU with 64 hyper-threads, 128\,GiB DDR4 memory, and a SOLIDIGM SSDPF2KX038T1 3.84\,TiB NVMe SSD, running Debian 9.13 with Linux kernel 5.4.56.

\begin{table}[!t]
\small
\setlength{\tabcolsep}{2pt}
\centering
\renewcommand{\arraystretch}{1.1}
\caption{Datasets used in evaluation.}
\label{tab:datasets}
\vspace{-9pt}
\begin{tabular}{|c|c|c|c|c|c|c|c|}
\hline
\textbf{Source} & \textbf{\#Vectors} & \textbf{Type} & \textbf{\#Dims} & \textbf{Size} & \textbf{\#Queries} \\
\hline
DecoupleVS100M & 109\,M & \texttt{FP32} & 128 & 53\,GiB & 10\,K  \\
\hline
SIFT100M \cite{sift} & 100\,M & \texttt{uint8} & 128 & 12\,GiB & 10\,K \\
\hline
SPACEV100M \cite{spacev} & 100\,M & \texttt{int8} & 100 & 9.4\,GiB & 29.3\,K \\
\hline
SIFT1B \cite{sift} & 1\,B & \texttt{uint8}  & 128 & 120\,GiB & 10\,K \\
\hline
SPACEV1B \cite{spacev} & 1.4\,B & \texttt{int8} & 100 & 131\,GiB & 29.3\,K \\
\hline
\end{tabular}
\vspace{-9pt}
\end{table}

\para{Datasets.}  Table~\ref{tab:datasets} shows the datasets. DecoupleVS100M is a proprietary dataset of 109\,M vectors dumped from a production-scale partition of a commercial search engine that indexes hundreds of billions of vectors. SIFT100M, SPACEV100M, SIFT1B, and SPACEV1B are public datasets; SIFT100M and SPACEV100M are the first 100\,M vectors of SIFT1B and SPACEV1B, respectively. SIFT100M and SIFT1B are derived from image descriptors; SPACEV100M and SPACEV1B are derived from web-search embeddings from Microsoft Bing.

\para{Baselines.} We compare \sysname against different disk-resident ANNS systems using their open-source prototypes.
\begin{itemize}[leftmargin=*]
\item 
{\em DiskANN} \cite{jayaram19}: a popular graph-based ANNS system employing best-first search with PQ-compressed vectors cached in memory.  
\item {\em PipeANN} \cite{guo25}: an extension of DiskANN that overlaps disk I/Os and distance computations during graph traversal to reduce search latency. 
\item
{\em FreshDiskANN} \cite{singh21}: an extension of DiskANN supporting streaming updates via a buffered update strategy.  
\item 
{\em OdinANN} \cite{guo26}: an extension of PipeANN supporting in-place vector updates to reduce latency and memory usage. 
\item 
{\em SPANN} \cite{chen21}: an on-disk cluster-based ANNS system that partitions vectors into clusters, connects centroids using an in-memory graph \cite{toussaint80}, and stores per-centroid posting lists on disk to reduce search I/Os. 
\end{itemize}

We build the graph indexes for PipeANN and \sysname using DiskANN's index-construction algorithm. We then decouple the storage and apply component-specific compression for \sysname atop the resulting DiskANN index. For 100\,M-scale datasets, index construction takes roughly 10~hours, whereas \sysname's compression and layout transformation complete in about 5~minutes, which is negligible compared to index construction.



\para{Default settings.} For DecoupleVS100M, SIFT100M, and SPACEV100M, the graph index uses $R=96$ outgoing edges per node and a candidate list size of $L_{b}=100$.  For SIFT1B and SPACEV1B, we increase to $R=128$ edges and $L_{b}=200$ to maintain high recall.  For SPANN, we use its default configurations.  We set the beam width $W=4$ for DiskANN, PipeANN, and \sysname, following DiskANN's default \cite{jayaram19}.  For fair comparison, we apply the same dynamic cache management (\S\ref{subsec:search}) to all systems, with a cache capacity of 1\,M entries (1\% and 0.1\% for 100\,M and 1\,B datasets, respectively).  For \sysname, we set the uncompressed chunk size $C=4$\,MiB, the uncompressed segment size to 512\,MiB, and the re-ranking batch size $B=10$.  We measure search accuracy by {\em recall@10}, the fraction of queries whose true nearest neighbors appear in the top-10 results ($K=10$).



\begin{figure}[!t]
\centering
\setlength{\tabcolsep}{0pt}
\begin{tabular}{cc}
\multicolumn{2}{c}{\includegraphics[height=24pt]{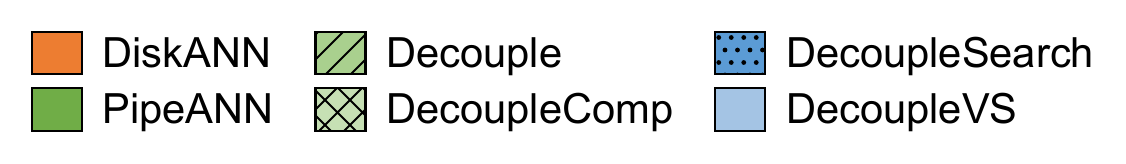}} \\
\multicolumn{2}{c}{\vspace{-16pt}} \\ 
\includegraphics[width=0.49\linewidth]{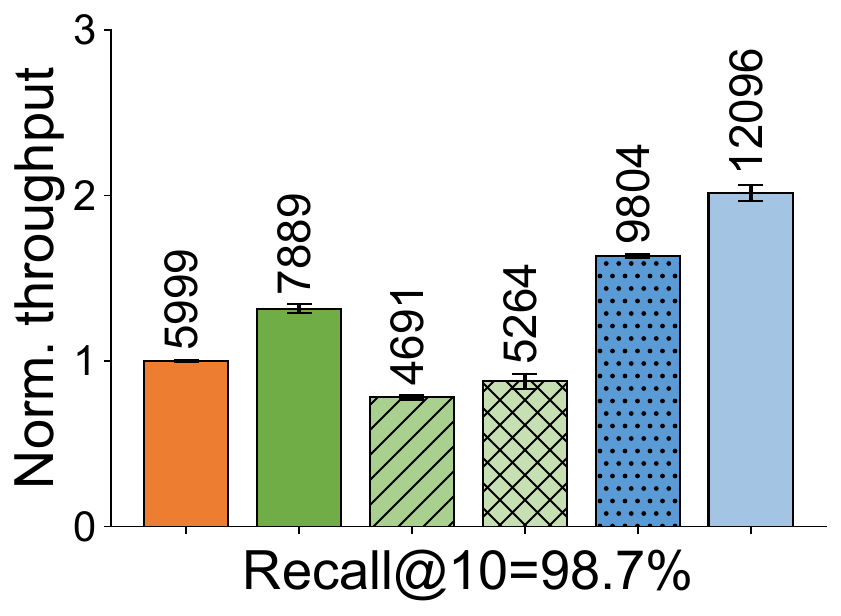} 
&
\includegraphics[width=0.49\linewidth]{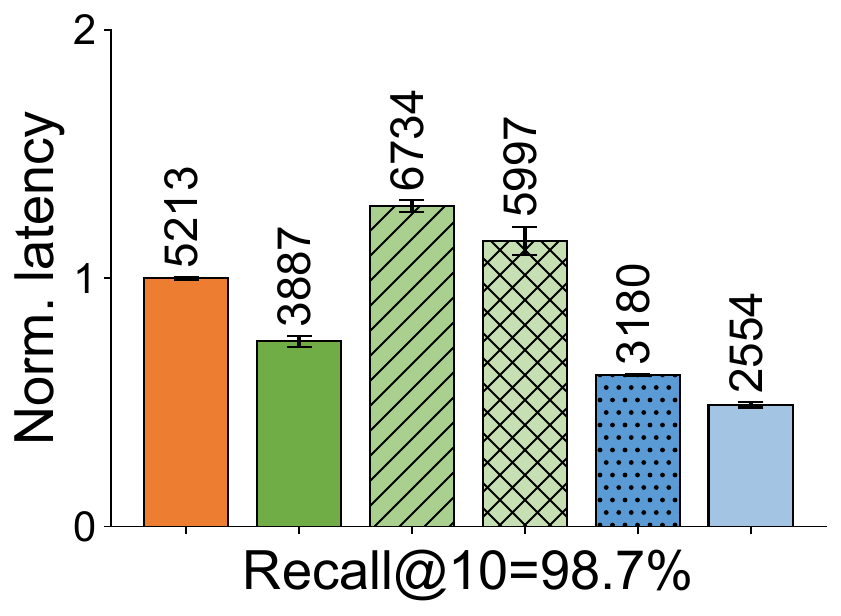}
\vspace{-3pt}\\
\parbox[t]{1.5in}{\centering \small (a) Search throughput} &
\parbox[t]{1in}{\centering \small (b) Search latency}
\end{tabular}
\vspace{-9pt}
\caption{Exp\#1 (Effectiveness of each component). Results are normalized to DiskANN. Numbers above bars are search throughput (QPS) and average search latency (ms).}
\label{fig:microbench}
\vspace{-6pt}
\end{figure}

\subsection{Microbenchmarks}
\label{subsec:micro}

\para{Exp\#1 (Effectiveness of each component).} We evaluate the contribution of \sysname's individual components to search performance by comparing throughput (queries-per-second (QPS)) and average latency (ms) across six configurations: (i) DiskANN, (ii) PipeANN (DiskANN with pipelined I/O), (iii) Decouple (i.e., PipeANN with the decoupled storage layout), (iv) DecoupleComp (i.e., Decouple with tailored lossless compression and hierarchical storage layouts, \S\ref{subsec:layout}), (v) DecoupleSearch (Decouple with latency-aware vector search, \S\ref{subsec:search}, but compression disabled), and (vi) \sysname. Figure~\ref{fig:microbench} shows throughput and average latency on SIFT100M, with all configurations targeting recall@10=98.7\%.  Overall, \sysname achieves the best performance, with a 2.02$\times$ throughput gain and 51.0\% latency reduction over DiskANN.  We explain these gains by tracing through the configuration stack. PipeANN improves throughput by 31.5\% over DiskANN via its I/O-overlapping strategy.  Adding the decoupled layout, Decouple degrades throughput by 21.8\% and 40.5\% over DiskANN and PipeANN, respectively, due to increased I/O overhead from storage separation. DecoupleComp improves 12.2\% throughput over Decouple by compressing the auxiliary index, allowing more metadata to reside in the LRU cache and reducing disk I/Os. DecoupleSearch, which adds latency-aware vector search without compression, achieves an 86.3\% throughput gain over Decouple, as adaptive vector prefetching removes full-precision vector I/Os from the critical search path and reduces latency by 39.0\% over DiskANN. Finally, \sysname enables both compression and latency-aware search, yielding a further 23.4\% throughput gain over DecoupleSearch by keeping more compressed index metadata in cache.

\subsection{Macrobenchmarks}
\label{subsec:overall}


We study \sysname's storage and performance efficiencies. \sysname achieves storage savings and search performance improvements through two independent mechanisms, making them {\em complementary} rather than a trade-off. Its component-aware lossless compression preserves data fidelity, with negligible decompression overhead (\S\ref{subsec:exp_analysis}). Its search performance improvements arise independently from latency-aware vector search (\S\ref{subsec:search}), which removes full-precision vector I/Os from the graph traversal critical path and reduces read amplification regardless of whether compression is used. Thus, \sysname consistently achieves lower storage overhead and higher search performance at similar recall across our evaluated workloads.

\begin{figure}[!t]
\centering
\setlength{\tabcolsep}{0pt}
\begin{tabular}{cc}
\multicolumn{2}{c}{\includegraphics[height=24pt]{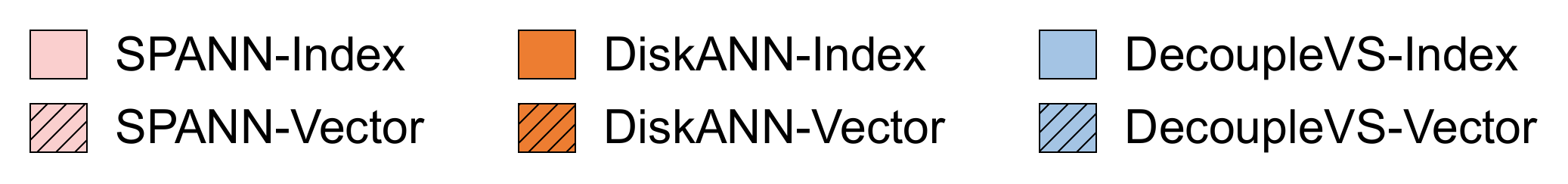}} \\
\multicolumn{2}{c}{\vspace{-16pt}} \\ 
\includegraphics[width=0.64\linewidth]{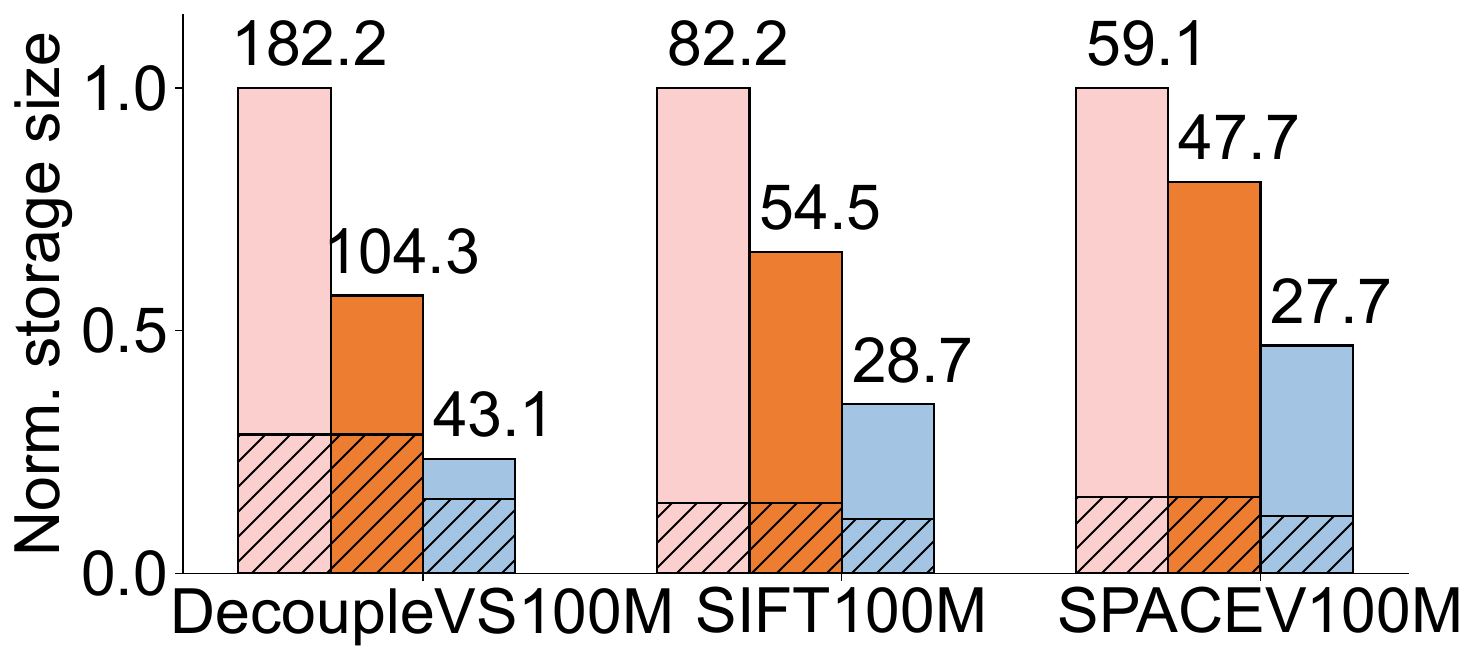} 
&
\includegraphics[width=0.36\linewidth]{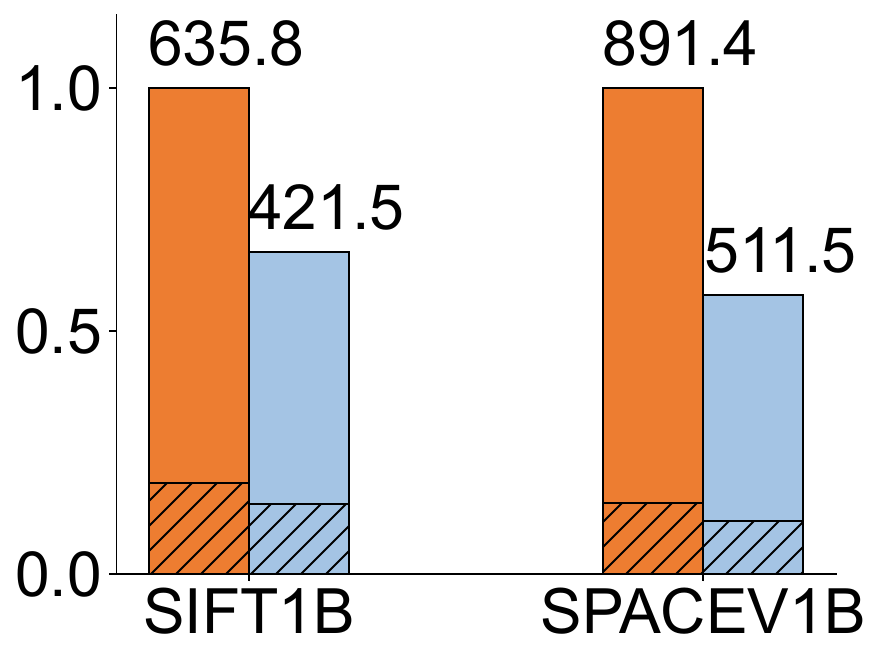}
\vspace{-3pt}\\
\parbox[t]{1in}{\centering \small (a) 100\,M-scale} &
\parbox[t]{1in}{\centering \small (b) 1\,B-scale}
\end{tabular}
\vspace{-9pt}
\caption{Exp\#2 (Storage savings). Sizes are normalized to SPANN for 100\,M-scale datasets and to DiskANN for 1\,B-scale datasets.  Numbers above bars indicate absolute storage sizes (GiB).}
\label{fig:space_efficiency}
\vspace{-6pt}
\end{figure}


\para{Exp\#2 (Storage savings).} We compare the storage sizes of \sysname against DiskANN and SPANN.  We omit PipeANN, which shares DiskANN's on-disk storage layout and hence identical storage footprint, and omit SPANN from billion-scale comparisons due to excessive memory usage during auxiliary index construction \cite{spannissue,guo25}.

Figure~\ref{fig:space_efficiency}(a) shows that \sysname achieves significant storage savings on 100\,M-scale datasets: 76.4\%, 65.1\%, and 53.1\% over SPANN, and 58.7\%, 47.4\%, and 41.9\% over DiskANN, on DecoupleVS100M, SIFT100M, and SPACEV100M, respectively.  SPANN's largest footprint stems from replicating each vector up to 8$\times$ across clusters \cite{chen21}. We further examine the per-component storage breakdown for DiskANN and \sysname. For vector data, \sysname reduces storage by 46.4\%, 23.8\%, and 25.2\% over raw vector sizes on DecoupleVS100M, SIFT100M, and SPACEV100M, respectively. Applying Huffman coding alone achieves only 38.5\% compression on DecoupleVS100M. This shows that XOR-delta encoding, when combined with Huffman coding, yields additional savings (\S\ref{subsec:compression}).  On SIFT100M and SPACEV100M, however, delta encoding provides no benefit as 8-bit quantization has already increased entropy, leaving limited headroom for further lossless compression. \sysname nonetheless achieves notable reductions on these datasets through Huffman coding alone.

Storage reductions in auxiliary index metadata stem from two sources: (i) DiskANN's page-aligned fixed-size entries incur internal fragmentation when vector and metadata sizes do not fit within 4 KiB boundaries. Decoupling mitigates this fragmentation, reducing the auxiliary index size by 19.5\%, 14.1\%, and 5.92\% on DecoupleVS100M, SIFT100M, and SPACEV100M, respectively, compared to DiskANN. (ii) Elias-Fano encoding further reduces the auxiliary index size by 51.5\%, 40.0\%, and 39.9\%, respectively. Together, these optimizations reduce the auxiliary index size by 71.0\%, 54.1\%, and 45.8\% compared to DiskANN on the three datasets.


Figure~\ref{fig:space_efficiency}(b) confirms that these gains persist at billion scale: \sysname reduces total storage by 33.7\% on SIFT1B and 42.6\% on SPACEV1B compared to DiskANN.


\begin{figure*}[!t]
\centering
\setlength{\tabcolsep}{0pt}
\begin{tabular}{ccccc}
\multicolumn{5}{c}{\includegraphics[height=16pt]{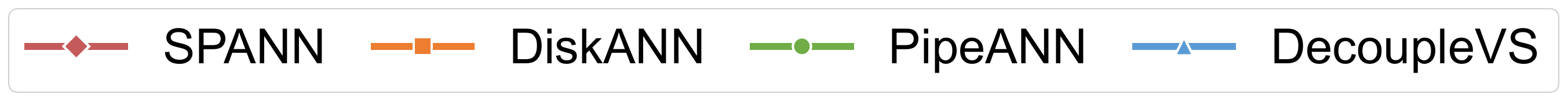}} \\
\multicolumn{5}{c}{\vspace{-16pt}} \\ 
\includegraphics[width=0.21\linewidth]{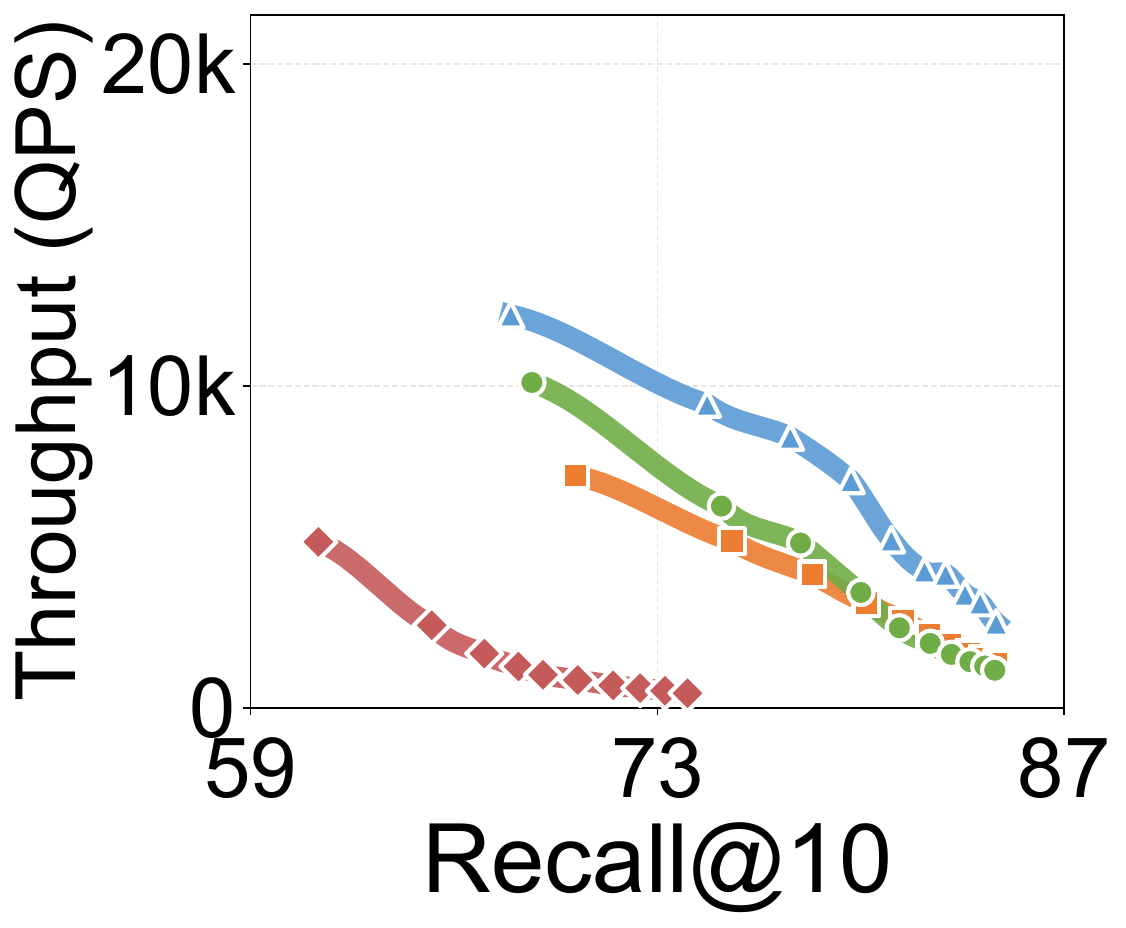} &
\includegraphics[width=0.2\linewidth]{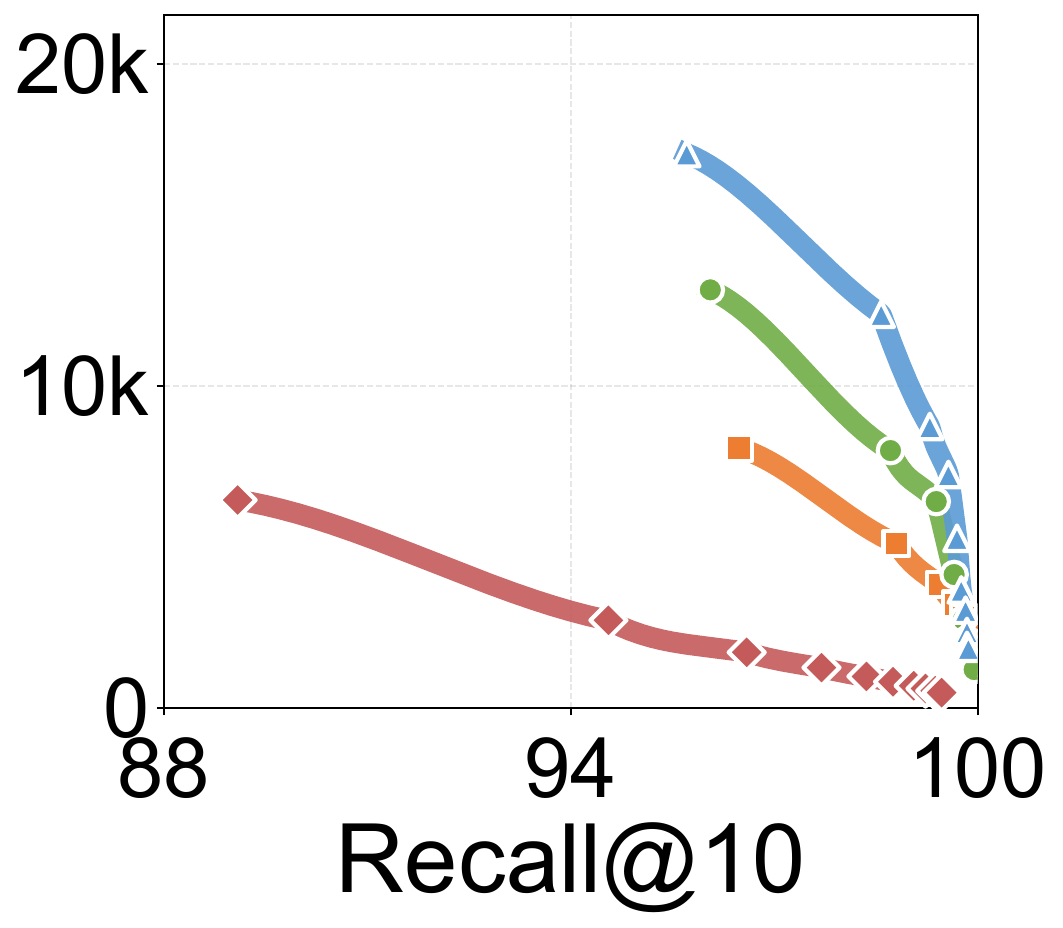} &
\includegraphics[width=0.2\linewidth]{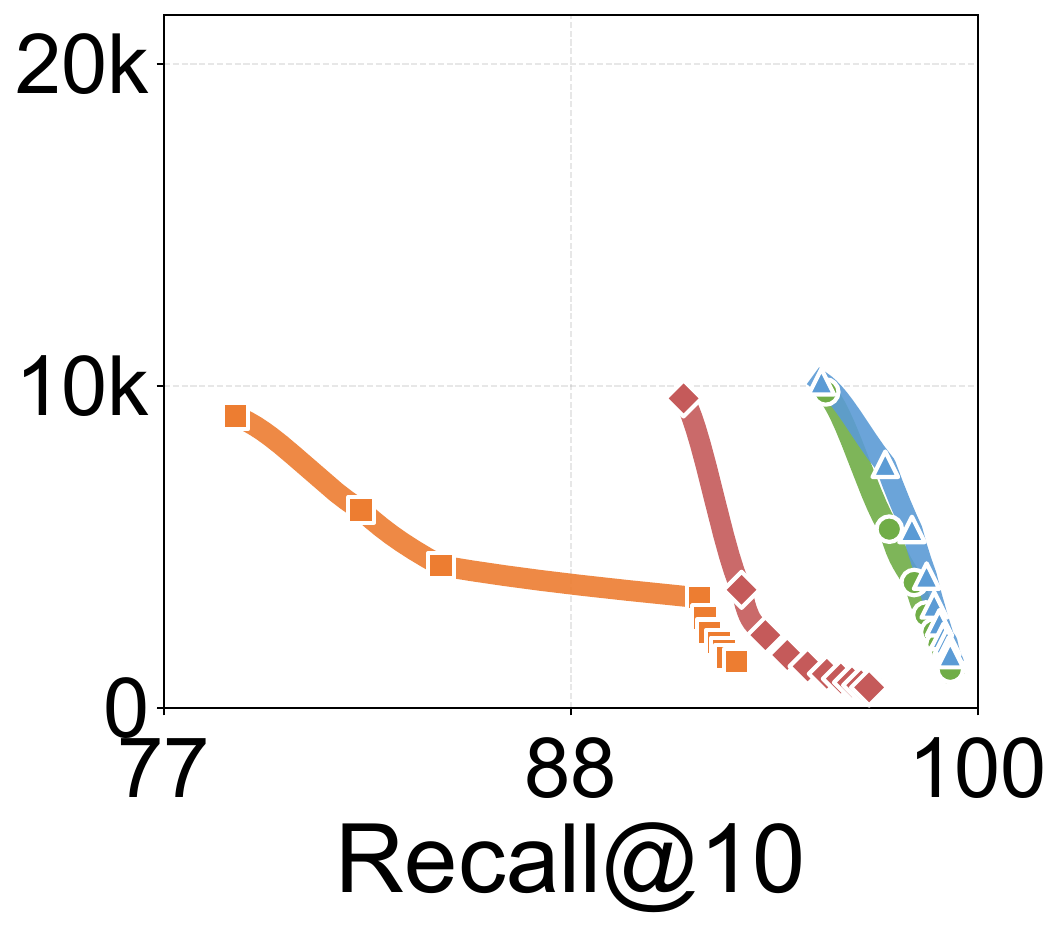} &
\includegraphics[width=0.2\linewidth]{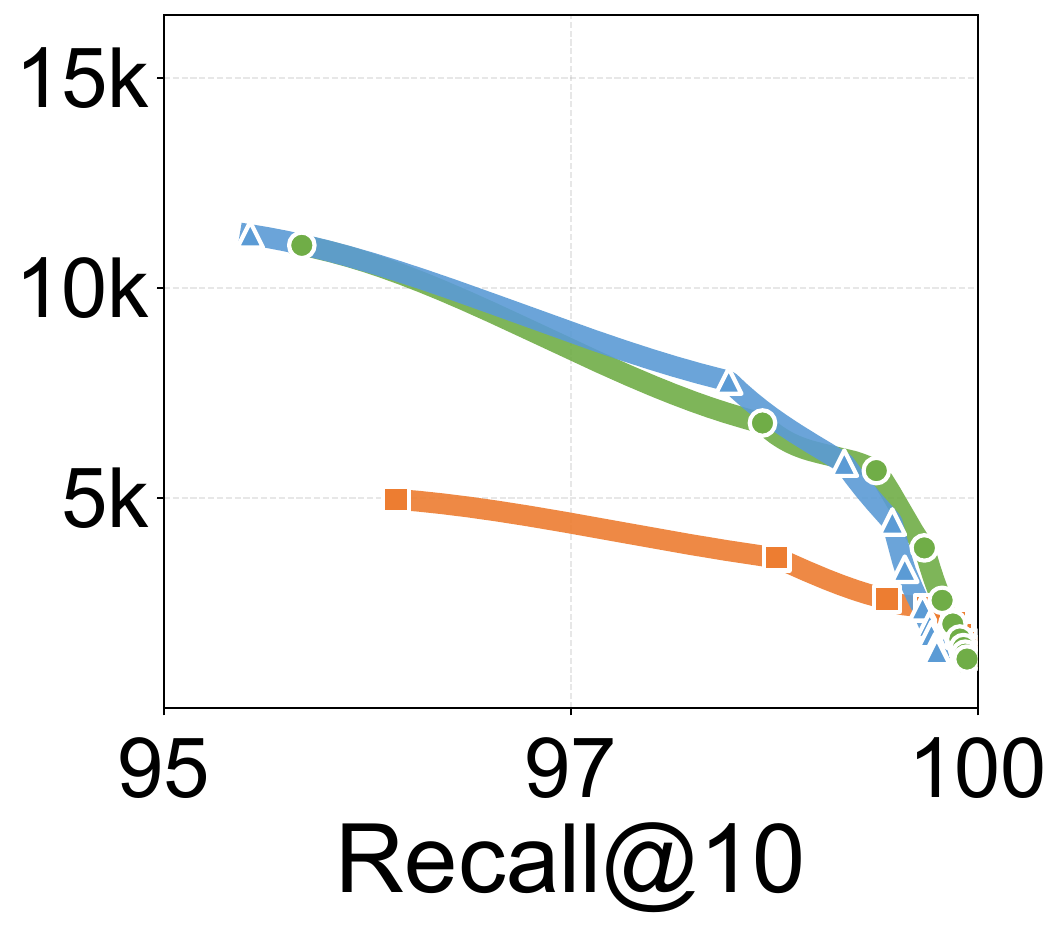} &
\includegraphics[width=0.195\linewidth]{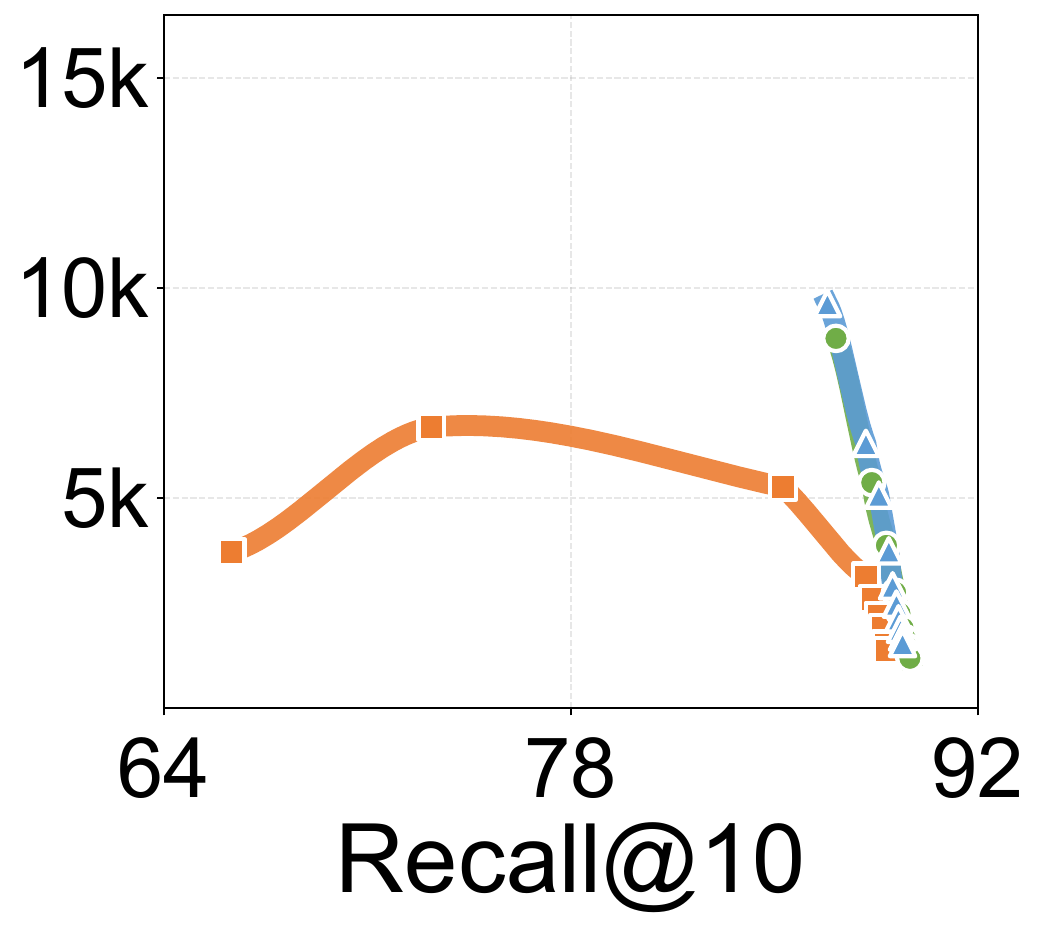} 
\vspace{-3pt}\\
\parbox[t]{1in}{\centering \small (a) DecoupleVS100M} &
\parbox[t]{1in}{\centering \small (b) SIFT100M} &
\parbox[t]{1in}{\centering \small (c) SPACEV100M} &
\parbox[t]{1in}{\centering \small (d) SIFT1B} &
\parbox[t]{1in}{\centering \small (e) SPACEV1B}
\end{tabular}
\vspace{-9pt}
\caption{Exp\#3 (Search throughput). Throughput (QPS) is plotted against Recall@10 (\%); higher is better. Points from left to right in each curve correspond to increasing candidate list sizes, which generally lead to higher recall.}
\label{fig:search_throughput}
\vspace{-6pt}
\end{figure*}

\begin{figure*}[!t]
\centering
\setlength{\tabcolsep}{0pt}
\begin{tabular}{ccccc}
\multicolumn{5}{c}{\includegraphics[height=16pt]{plot/evaluation/search_throughput/legend.pdf}} \\
\multicolumn{5}{c}{\vspace{-16pt}} \\ 
\includegraphics[width=0.21\linewidth]{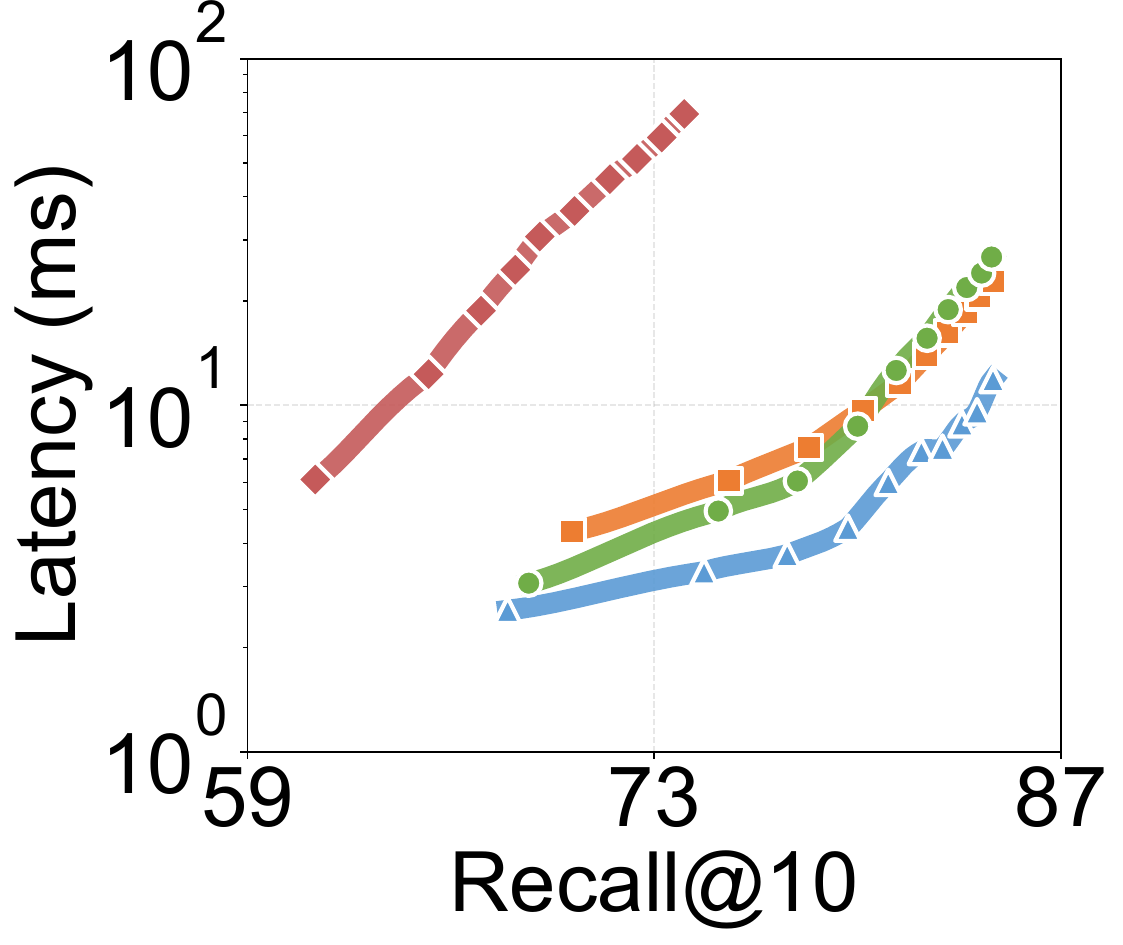} &
\includegraphics[width=0.2\linewidth]{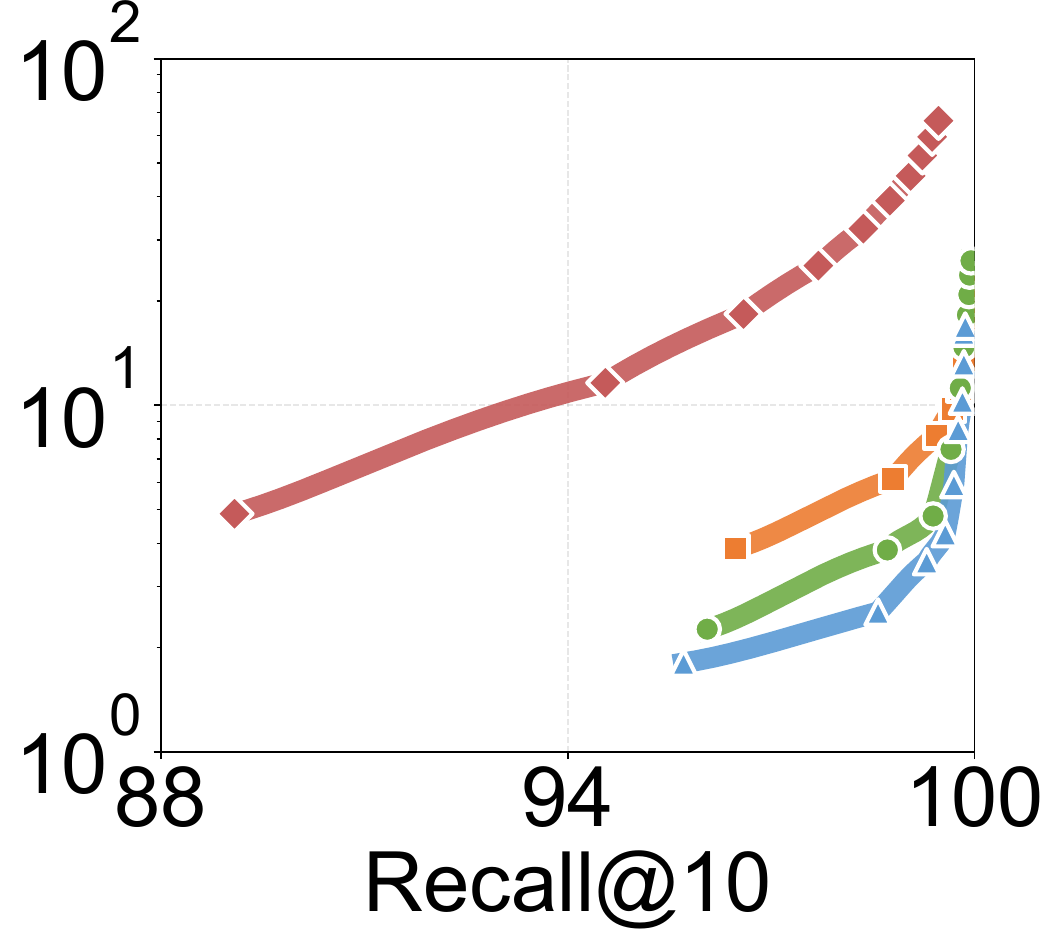} &
\includegraphics[width=0.2\linewidth]{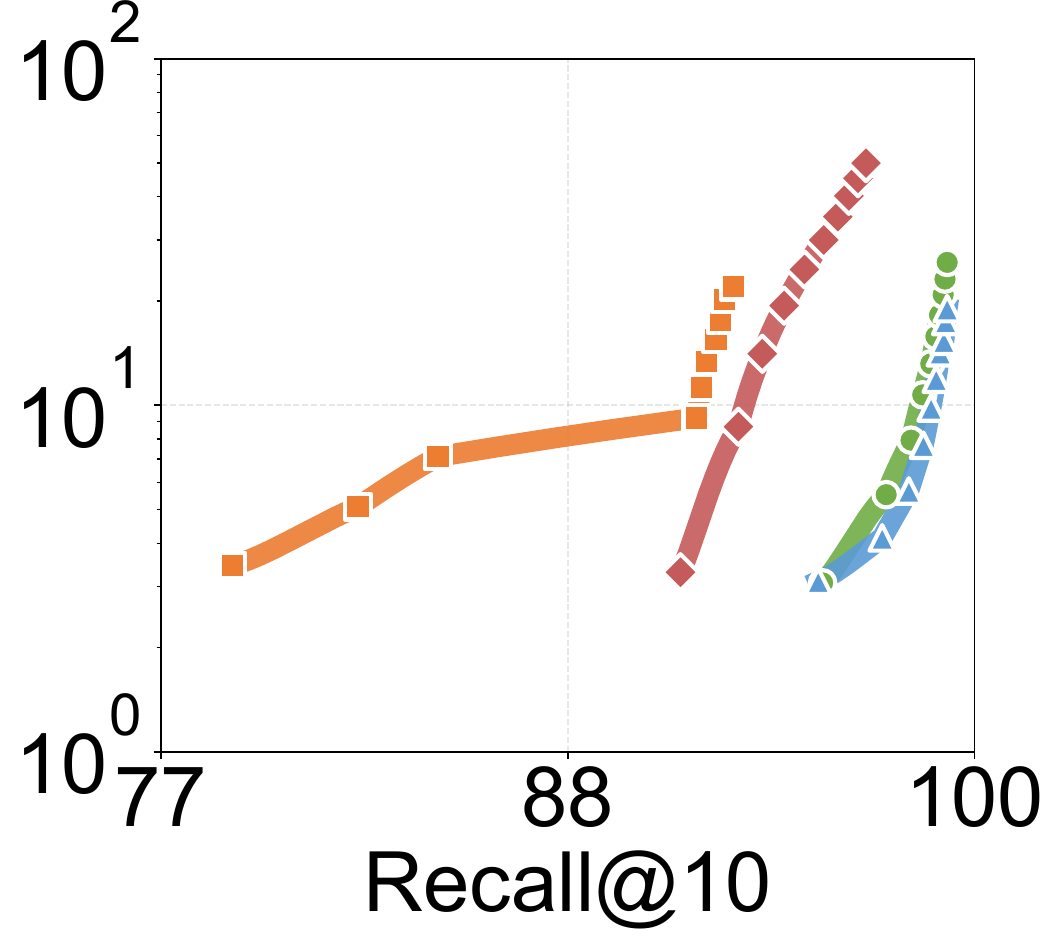} &
\includegraphics[width=0.2\linewidth]{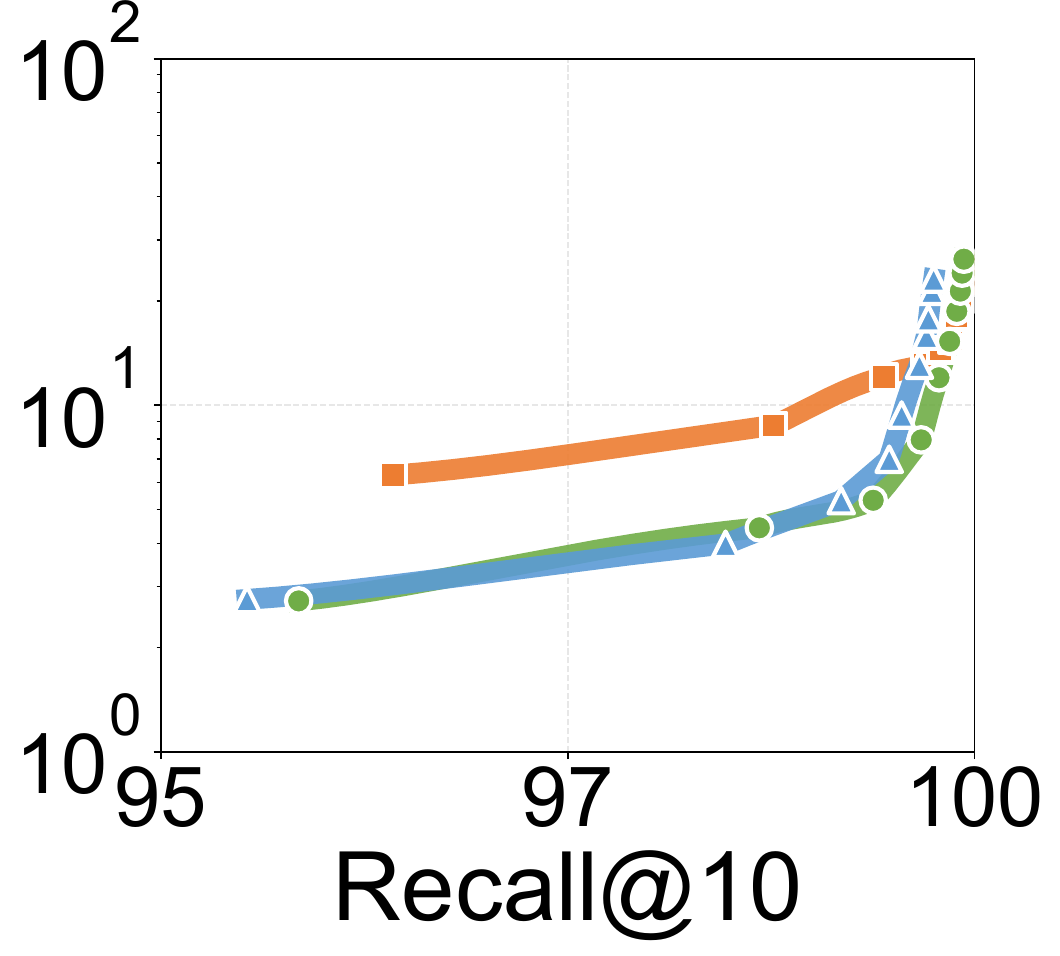} &
\includegraphics[width=0.195\linewidth]{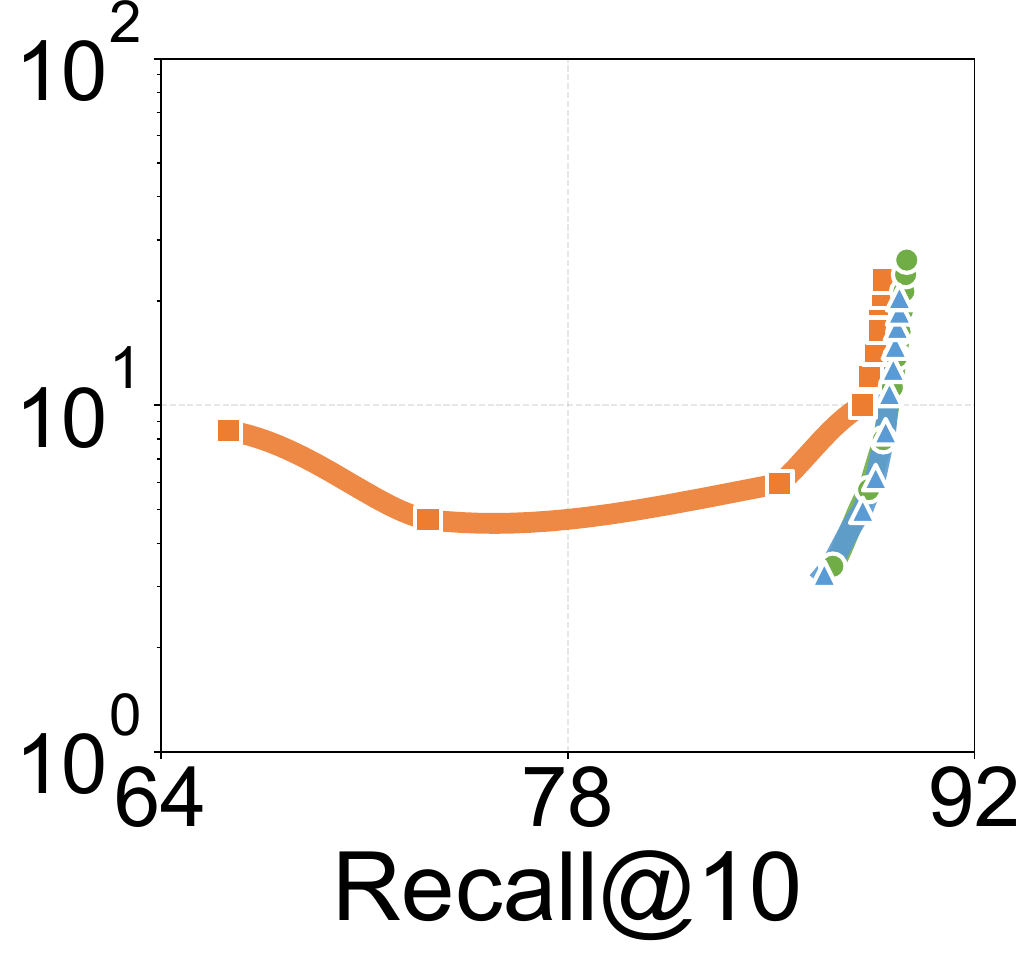}
\vspace{-3pt}\\
\parbox[t]{1in}{\centering \small (a) DecoupleVS100M} &
\parbox[t]{1in}{\centering \small (b) SIFT100M} &
\parbox[t]{1in}{\centering \small (c) SPACEV100M} &
\parbox[t]{1in}{\centering \small (d) SIFT1B} &
\parbox[t]{1in}{\centering \small (e) SPACEV1B}
\end{tabular}
\vspace{-9pt}
\caption{Exp\#4 (Search latency). Average latency (ms) is plotted against
Recall@10 (\%); lower is better.}
\label{fig:search_latency}
\vspace{-6pt}
\end{figure*}

\para{Exp\#3 (Search throughput).} We evaluate search throughput (QPS) of \sysname against SPANN, DiskANN, and PipeANN, varying the search candidate list size $L_{s}$ from 50 to 500 to obtain 10 accuracy levels for DiskANN and PipeANN (graph-based), and adjusting SPANN's internal result count within the same range.  We use 64 concurrent search threads to maximize throughput. 

Figures~\ref{fig:search_throughput}(a)-\ref{fig:search_throughput}(c) show search throughput on the three 100\,M-scale datasets.  \sysname achieves the highest throughput across all baselines at every accuracy level. On DecoupleVS100M at recall@10 $=83.1\% \pm 0.1\%$, \sysname achieves 1.88$\times$ and 2.13$\times$ throughput gains over DiskANN and PipeANN, respectively (Figure~\ref{fig:search_throughput}(a)).  On SIFT100M at recall@10 $= 98.8\% \pm 0.1\%$, \sysname achieves 2.39$\times$ and 1.53$\times$ throughput gains, respectively (Figure~\ref{fig:search_throughput}(b)). These improvements stem from higher cache space utilization and the differentiated I/O strategy that removes vector fetching from the critical search path (\S\ref{subsec:search}). Graph-based systems generally outperform SPANN, except DiskANN on SPACEV100M, where SPANN performs slightly better at high accuracy. Both PipeANN and \sysname surpass SPANN's throughput and peak accuracy: they benefit from PipeANN's enlarged search space and overlapped I/O execution, which together yield a better accuracy-throughput frontier. 


Figures~\ref{fig:search_throughput}(d)-\ref{fig:search_throughput}(e) show that \sysname maintains its advantage
on billion-scale datasets: it achieves 2.17$\times$ and 1.15$\times$ throughput gains over DiskANN and PipeANN on SIFT1B at recall@10 $=98.7\% \pm 0.1\%$, respectively (Figure~\ref{fig:search_throughput}(d)), and 2.0$\times$ and 1.17$\times$ gains on SPACEV1B at recall@10 $=88.8\% \pm 0.1\%$, respectively (Figure~\ref{fig:search_throughput}(e)).
  
\para{Exp\#4 (Search latency).}  We evaluate the average search latency (ms), following the same setting in Exp\#3. Figure~\ref{fig:search_latency} shows that the latency results are consistent with the throughput findings. On DecoupleVS100M at recall@10 $=83.1\% \pm 0.1\%$, \sysname reduces latency by 47.2\% over DiskANN and 52.9\% over PipeANN (Figure~\ref{fig:search_latency}(a)). On SIFT1B at recall@10 $= 98.7\% \pm 0.1\%$, \sysname reduces latency by 54.6\% over DiskANN and 11.0\% over PipeANN (Figure~\ref{fig:search_latency}(d)).  In Appendix (see the supplementary file), we show that \sysname reduces P99 tail latency by 51.9\% over both DiskANN and PipeANN on DecoupleVS100M at recall@10$=83.0\%\pm0.1\%$, confirming that its latency advantage holds at the tail.

\para{Exp\#5 (Concurrent search and update performance).} We compare \sysname against FreshDiskANN \cite{singh21}(buffered updates) and OdinANN \cite{guo26} (in-place updates). We use the best-first search for FreshDiskANN and PipeSearch \cite{guo25} for OdinANN. All systems are indexed on SIFT100M. As in \cite{singh21,guo26}, we simulate a streaming update workload by replacing 50\% of vectors over 10 iterations, each merging 5\% deletions and 5\% insertions. We launch 32 concurrent search threads and 32 update threads, and measure throughput, P50 and P99 latencies, recall, memory usage, and storage size throughout.

\begin{figure*}[!t]
\centering
\setlength{\tabcolsep}{0pt}
\begin{tabular}{ccc}
\multicolumn{3}{c}{\includegraphics[height=16pt]{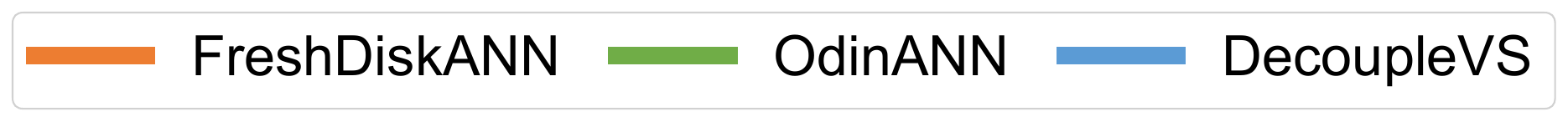}}
\vspace{-3pt}\\ 
\includegraphics[width=0.33\linewidth]{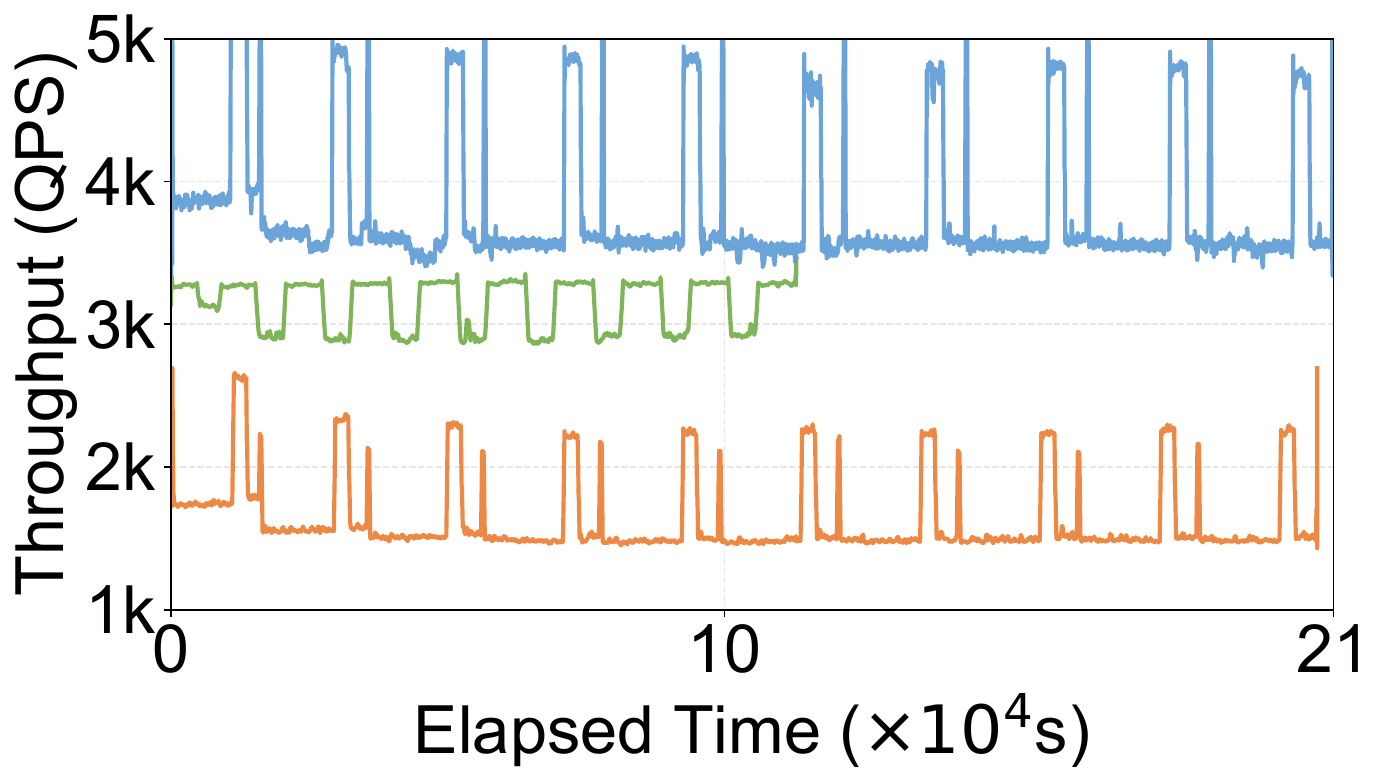} &
\includegraphics[width=0.33\linewidth]{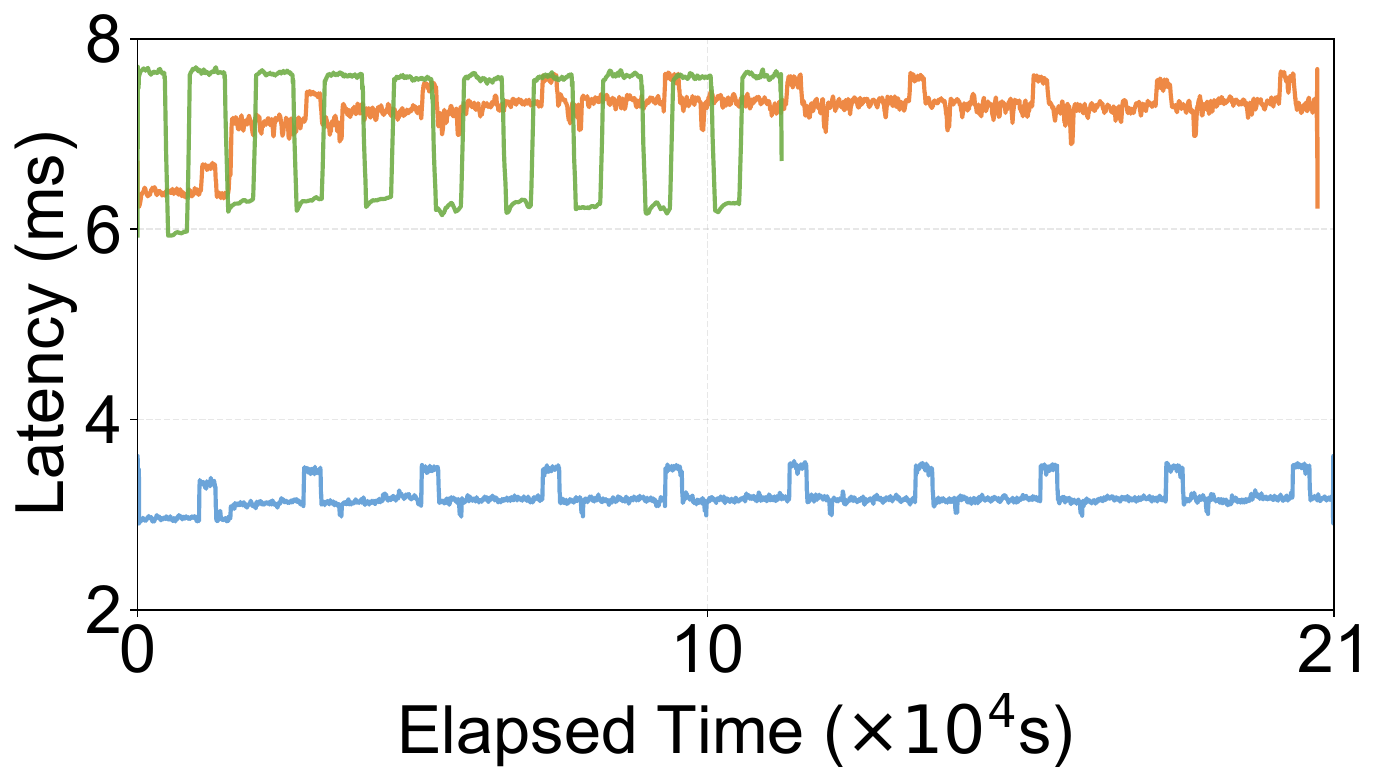} &
\includegraphics[width=0.33\linewidth]{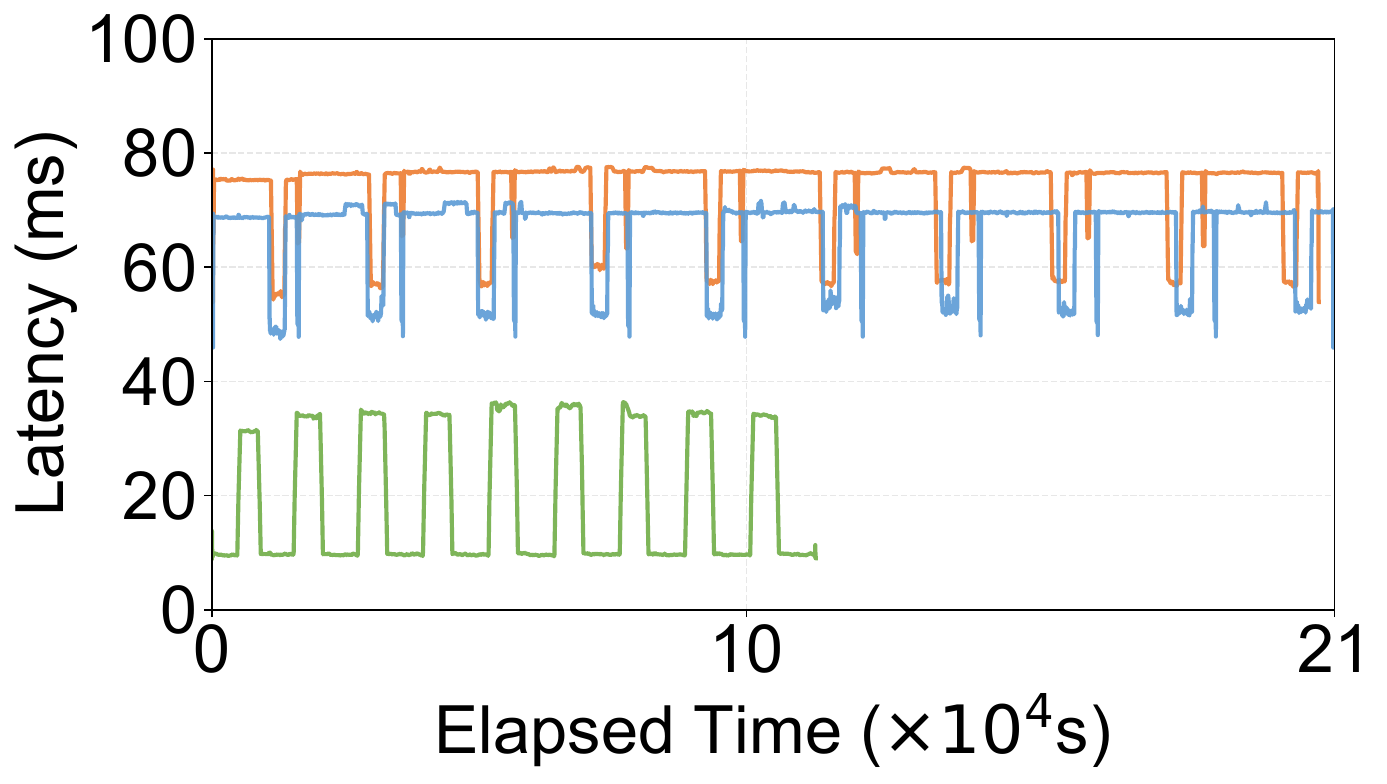} 
\vspace{-3pt}\\
\parbox[t]{2in}{\centering \small (a) Search throughput} &
\parbox[t]{1in}{\centering \small (b) P50 latency} &
\parbox[t]{1in}{\centering \small (c) P99 latency} 
\vspace{3pt}\\
\includegraphics[width=0.33\linewidth]{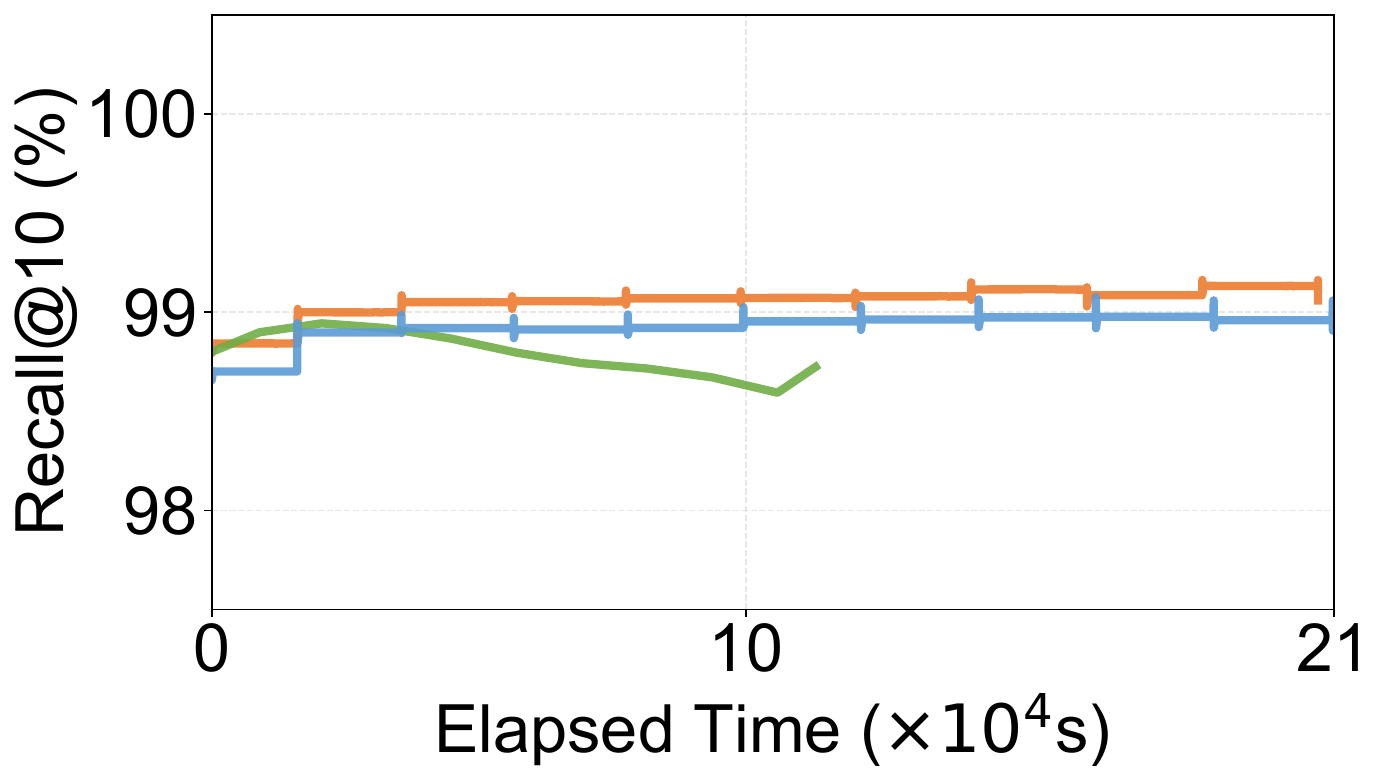} &
\includegraphics[width=0.33\linewidth]{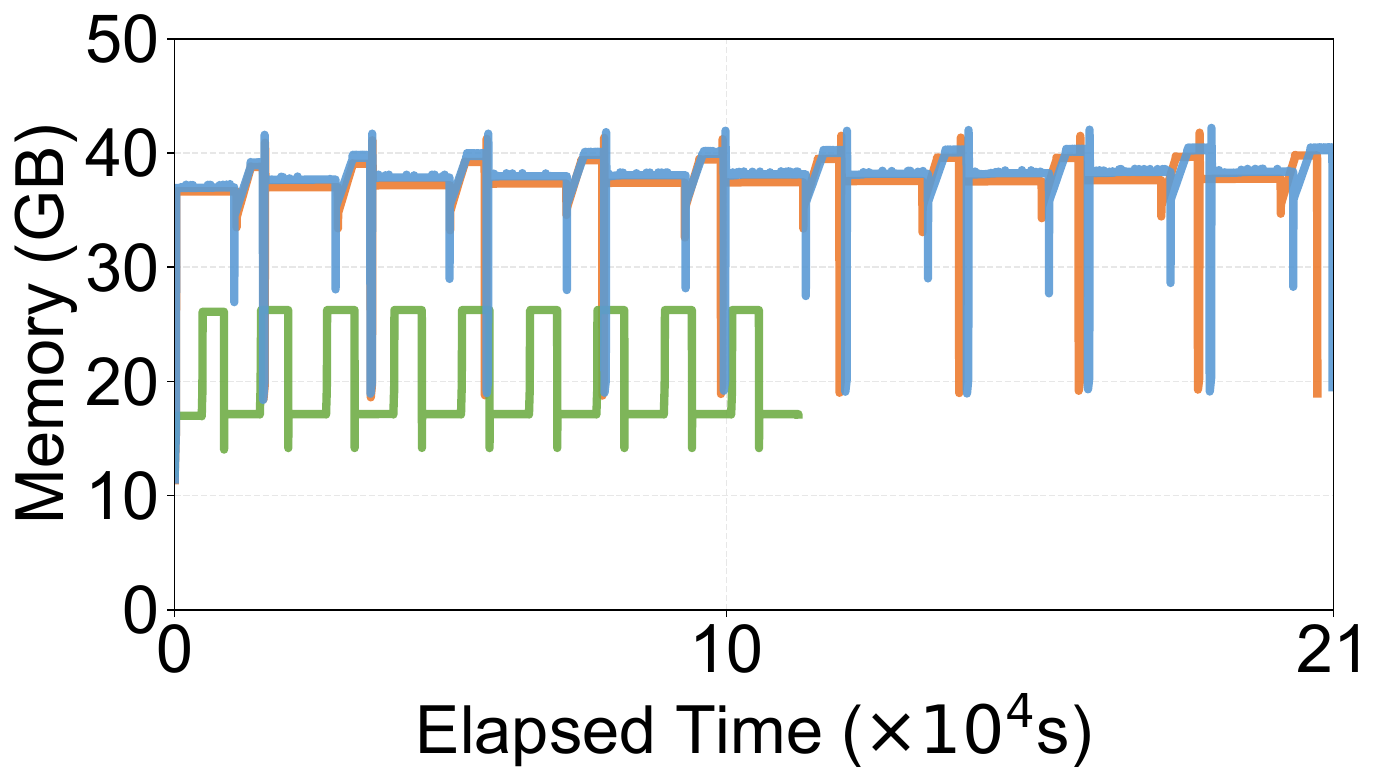} &
\includegraphics[width=0.33\linewidth]{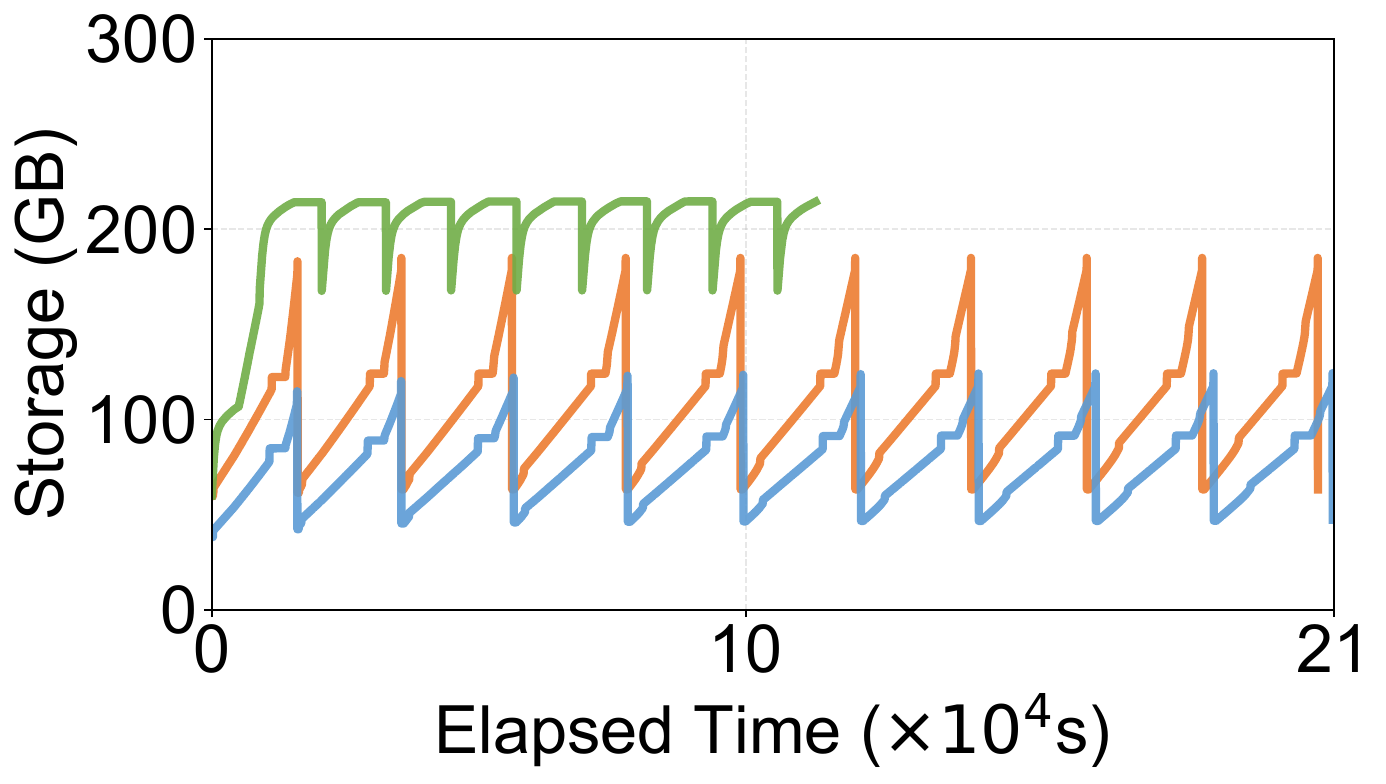} 
\vspace{-3pt}\\
\parbox[t]{1in}{\centering \small (d) Search accuracy} &
\parbox[t]{1in}{\centering \small (e) Memory usage} &
\parbox[t]{1in}{\centering \small (f) Storage size}
\end{tabular}
\vspace{-9pt}
\caption{Exp\#5 (Concurrent search and updates). Performance is measured on SIFT100M.}
\label{fig:update_performance}
\vspace{-6pt}
\end{figure*}

Figures~\ref{fig:update_performance}(a)-\ref{fig:update_performance}(b) show that \sysname consistently achieves higher throughput and lower P50 latency than both baselines, with average throughput gains of 2.31$\times$ and 1.21$\times$ over FreshDiskANN and OdinANN, and average P50 latency reductions of 55.5\% and 55.4\%, respectively. These gains stem from \sysname's latency-aware search strategy (\S\ref{subsec:search}), which mitigates I/O interference from concurrent updates.  OdinANN, however, achieves the lowest P99 latency, reducing it by 77.5\% and 75.0\% compared to FreshDiskANN and \sysname, respectively (Figure~\ref{fig:update_performance}(c)), as its delta neighbor pruning substantially reduces per-update computation, alleviating tail latency. Delta neighbor pruning is orthogonal to \sysname's design and can be incorporated to further reduce P99 latency. In terms of recall, \sysname matches FreshDiskANN within 0.13\% on average, while OdinANN incurs a 0.26\% average loss (Figure~\ref{fig:update_performance}(d)).

Figures~\ref{fig:update_performance}(e)-\ref{fig:update_performance}(f) show space efficiency during updates. 
\sysname's peak memory usage is only 0.98\% higher than FreshDiskANN's, attributed to: (i) in-memory compression metadata (28\,MiB on SIFT100M), (ii) compression and decompression buffers during updates, and (iii) larger per-batch entry counts under the compressed layout.  
For SIFT100M, the compression metadata totals 28.5\,MiB (including 19.6\,MiB for the sparse in-memory index, 8.84\,MiB for chunk metadata, and 30\,KiB for Huffman codebooks), accounting for only 0.095\% of storage size. OdinANN reduces memory usage by 37.2\% over FreshDiskANN and 37.8\% over \sysname by avoiding in-memory index buffering.  At the storage layer, \sysname reduces footprint by 26.9\% over FreshDiskANN and 61.8\% over OdinANN; OdinANN's largest footprint is due to the additional on-disk structures required for in-place updates. The stable storage size across merge iterations confirms that GC successfully reclaims stale vector space.

In summary, \sysname achieves the best search performance during concurrent updates with storage overhead comparable to FreshDiskANN. These results reflect an inherent trade-off between update strategies: buffered updates (in FreshDiskANN and \sysname) preserve storage efficiency and search performance at the cost of higher memory usage, while in-place updates (in OdinANN) reduce memory at the cost of a larger storage footprint. Notably, both strategies are orthogonal to \sysname's decoupled storage layout and can be combined with it to achieve high space efficiency.



\subsection{System Analysis}
\label{subsec:exp_analysis}

We analyze SIFT100M to provide detailed breakdowns of search and update performance. We also study the impact of compression algorithms using the 100\,M-scale datasets. 

\para{Exp\#6 (Search performance breakdown).} We decompose the per-search-query resource usage of DiskANN, PipeANN, and \sysname at $L_s=100$ into I/O and CPU components: the I/O usage component include: (i) graph cache hit count, (ii) graph traversal disk I/O count, (iii) vector fetch disk I/O count, and (iv) total disk I/O time; the CPU metrics include: (i) graph decompression time, (ii) PQ distance calculation time, (iii) vector decompression time, and (iv) re-ranking distance calculation time. Each metric is averaged across all queries. As 64 search threads run concurrently, the wall-clock time per query exceeds the sum of individual step times due to inter-thread contention.

Table~\ref{tab:breakdown} shows the breakdown. For I/O usage, \sysname is highly cache-efficient, increasing graph cache hit count by 46.8\% over DiskANN and 120.6\% over PipeANN. PipeANN has the lowest cache hit count since its enlarged search space evicts entries more aggressively, resulting in the highest total disk I/Os (67.9 versus DiskANN's 64.5 and \sysname's 44.1) and the highest total disk I/O time. PipeANN partially recovers by overlapping I/O with computation, yielding lower overall latency than DiskANN despite more I/Os. For CPU usage, \sysname's decompression overhead, covering both neighbor lists and full-precision vectors, accounts for only 4.1\% of average query latency, confirming that compression does not impede search. DiskANN exhibits the highest CPU computation time because its blocking I/O model causes frequent context switches across threads, thereby increasing the number of vectors visited. \sysname's PQ calculation time is higher than PipeANN's due to higher CPU utilization from reduced I/O wait time. In summary, disk I/O is the dominant cost in all systems; \sysname reduces I/O time by 42.5\% over DiskANN and 48.9\% over PipeANN, confirming its effectiveness in reducing read amplification.

\begin{table}[!t]
\footnotesize
\setlength{\tabcolsep}{5pt}
\centering
\caption{Exp\#6 (Search performance breakdown). We show the average per-search-query breakdown latencies (ms), including 95\% confidence intervals.}
\label{tab:breakdown}
\vspace{-9pt}
\begin{threeparttable}
\renewcommand{\arraystretch}{1.2}
\begin{tabular}{|c|c|c|c|}
\hline
\textbf{Steps} & \textbf{DiskANN} & \textbf{PipeANN} & \textbf{\sysname} \\ 
\hline 
Total time & 5.67$\pm$0.60 & 3.97$\pm$0.04 & 2.48$\pm$0.14 \\
\hline
\multicolumn{4}{|c|}{\textbf{I/O usage}} \\
\hline
Graph cache hits & 66.3$\pm$0.01 & 44.1$\pm$0.02 & 97.3$\pm$0.08 \\
\hline 
Graph I/Os & 64.5$\pm$0.01 & 67.9$\pm$0.03 & 10.9$\pm$0.09 \\
\hline 
Vector I/Os & - & - & 33.2$\pm$0.03 \\
\hline
Total I/O time & 2.33$\pm$0.04 & 2.62$\pm$0.14 & 1.34$\pm$0.07\\
\hline
\hline
\multicolumn{4}{|c|}{\textbf{CPU usage}} \\
\hline
Graph decompression & - & - & 0.031$\pm$0.008 \\
\hline
PQ calculation & 0.64$\pm$0.02 & 0.60$\pm$0.07 & 0.44$\pm$0.04 \\
\hline
Vector decompression & - & - & 0.070$\pm$0.005 \\
\hline
Rerank calculation & 0.023$\pm$0.005 & 0.012$\pm$0.005 & 0.0028$\pm$0.0001\\
\hline
\end{tabular}
\begin{tablenotes}[para,flushleft]
\footnotesize
* DiskANN and PipeANN co-locate vector data with auxiliary index metadata, so no separate vector I/O step is counted.
\end{tablenotes}
\end{threeparttable}
\vspace{-6pt}
\end{table}

\para{Exp\#7 (Update performance breakdown).} We analyze the two most time-consuming update operations in \sysname: (i) Merge-Delete (merging deletion markers into the on-disk index) and (ii) Merge-Insert (merging newly inserted points into the on-disk index). Each component is decomposed into computation and disk I/O time, and the results are averaged over 10 iterations, with 95\% confidence intervals based on the Student's t-distribution. We compare FreshDiskANN, \sysname, and \sysname-NoGC (a \sysname variant with GC disabled to isolate the GC impact on update performance). 

\begin{figure}[!t]
\centering
\setlength{\tabcolsep}{0pt}
\begin{tabular}{@{\ }c@{\ }c}
\multicolumn{2}{c}{\includegraphics[height=16pt]{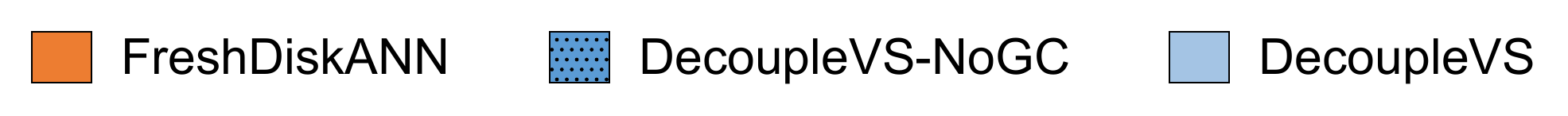}} 
\vspace{-6pt}\\
\includegraphics[width=0.49\linewidth]{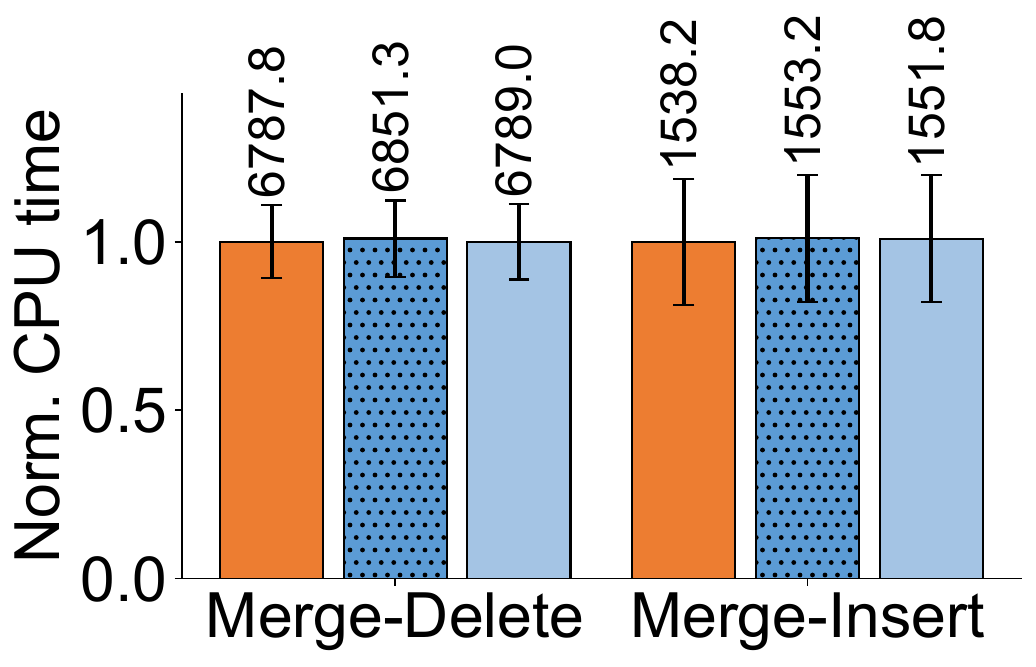} 
&
\includegraphics[width=0.49\linewidth]{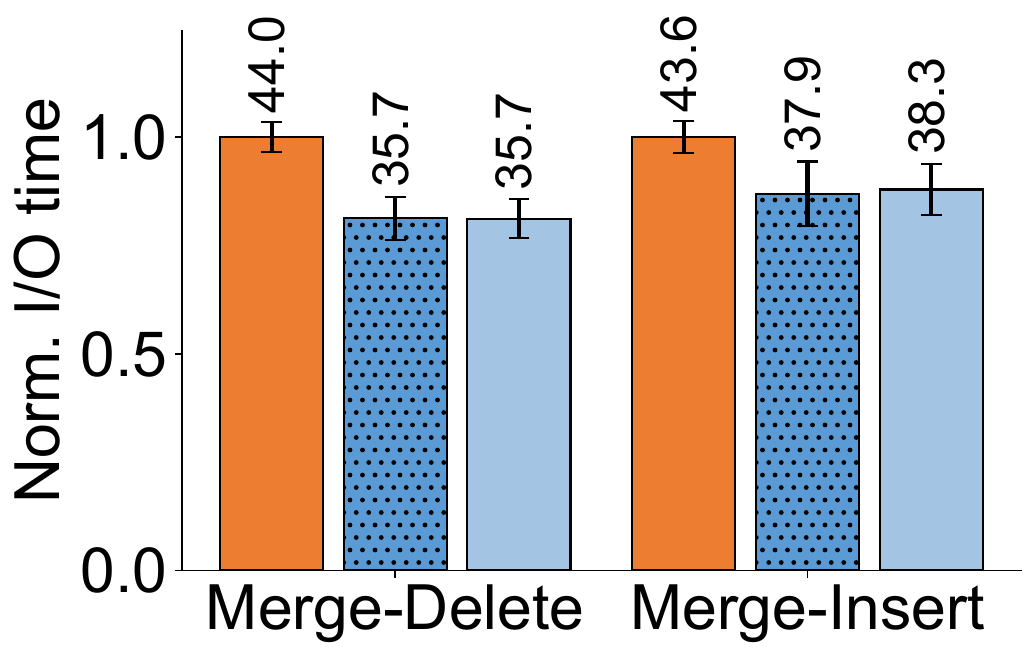}
\vspace{-3pt}\\
{\small (a) Computation time} & 
{\small (b) Disk I/O time}
\end{tabular}
\vspace{-12pt}
\caption{Exp\#7 (Update performance breakdown). Average computation and disk I/O time per update operation are normalized to DiskANN. Numbers above bars indicate absolute time in seconds.}
\label{fig:update_breakdown}
\vspace{-9pt}
\end{figure}

Figure~\ref{fig:update_breakdown} shows the results normalized to FreshDiskANN. The computation time is nearly identical across all three schemes: \sysname incurs only 0.018\% and 0.88\% increases over FreshDiskANN for Merge-Delete and Merge-Insert, respectively, attributed to compression and decompression.  \sysname reduces disk I/O time by 18.9\% for Merge-Delete and 12.2\% for Merge-Insert over FreshDiskANN, confirming that \sysname's decoupled layout reduces write amplification. \sysname and \sysname-NoGC differ by at most 0.9\% in computation time and 1.1\% in disk I/O time, both within the margin of error, indicating that GC imposes negligible overhead on foreground update operations.

\begin{figure}[!t]
\centering
\setlength{\tabcolsep}{0pt}
\begin{tabular}{cc}
\multicolumn{2}{c}{\includegraphics[height=14pt]{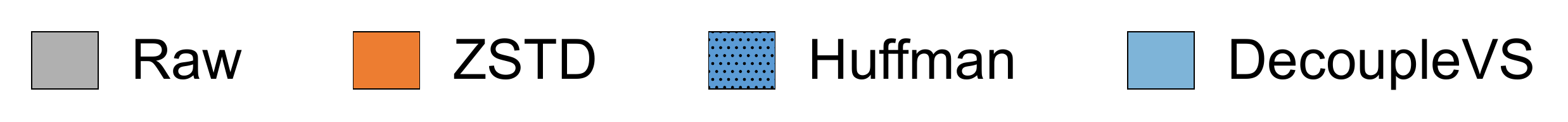}} \\
\multicolumn{2}{c}{\vspace{-16pt}} \\ 
\includegraphics[width=0.57\linewidth]{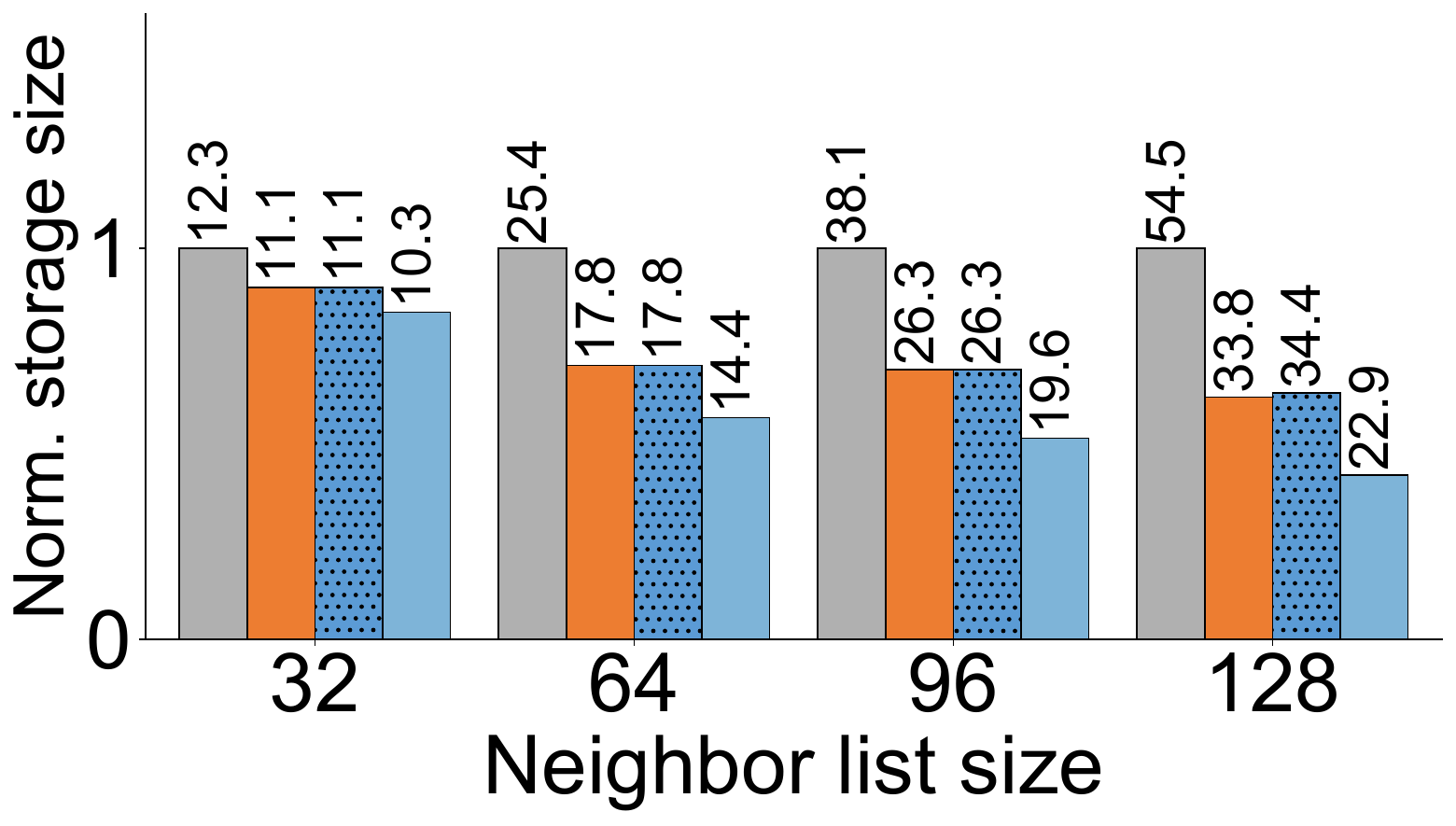} 
&
\includegraphics[width=0.43\linewidth]{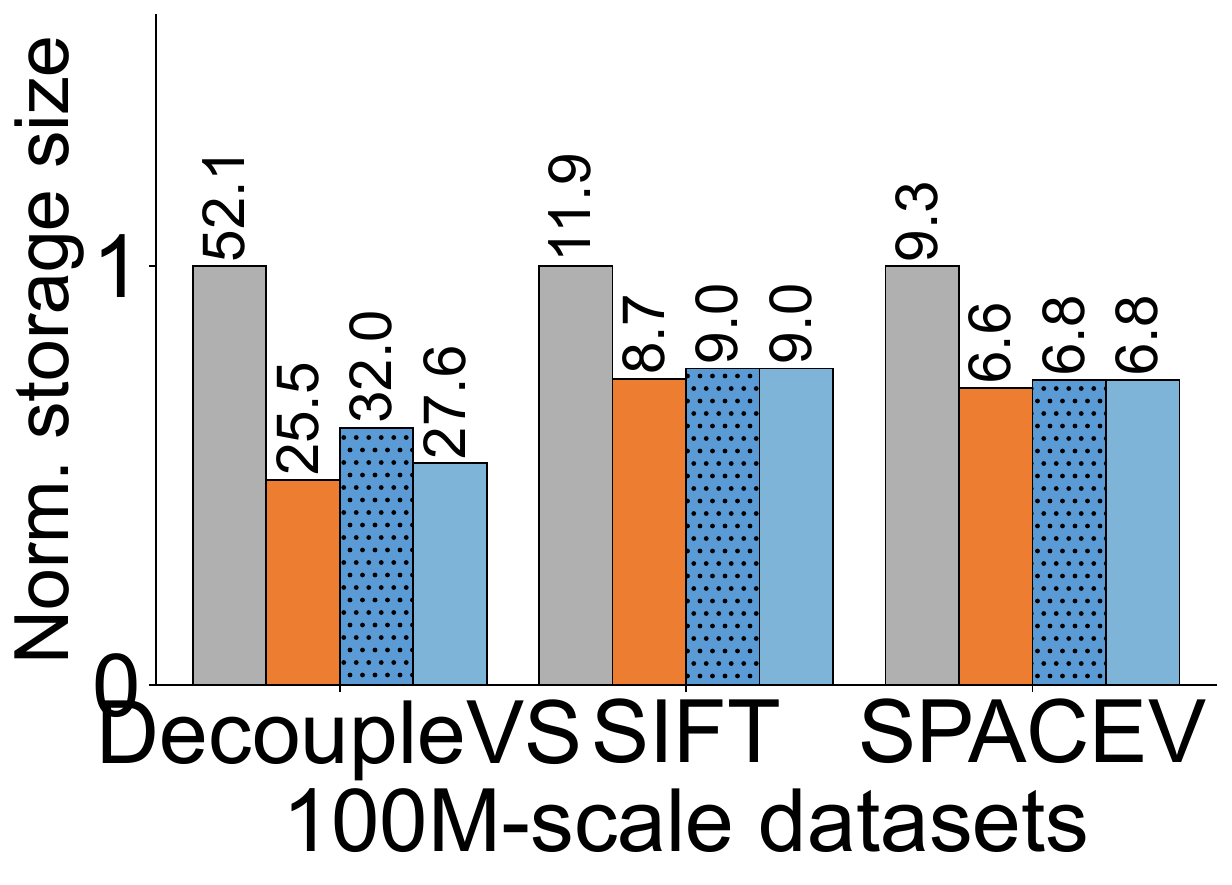}
\vspace{-6pt}
\\
\parbox[t]{2in}{\centering \small (a) Auxiliary index compression} &
\parbox[t]{1.5in}{\centering \small (b) Vector compression}
\end{tabular}
\vspace{-12pt}
\caption{Exp\#8 (Compression algorithm analysis). Sizes are normalized to the raw size, and numbers atop the bars indicate absolute storage sizes in GiB.}
\label{fig:compression_alg}
\vspace{-9pt}
\end{figure}

\para{Exp\#8 (Compression algorithm analysis).}
We compare \sysname against two general-purpose lossless compression techniques, ZSTD and Huffman, using the Zstandard library \cite{zstd}, which implements both algorithms. Figure~\ref{fig:compression_alg}(a) shows the storage reduction on the auxiliary index metadata as a function of neighbor list size $R$. \sysname consistently achieves the largest reduction, as Elias-Fano encoding is specifically designed for monotone integer sequences. At $R=96$, \sysname reduces index size by 48.6\%, versus 31.0\% for both ZSTD and Huffman. The advantage of \sysname grows with a larger $R$, since longer sorted integer lists become increasingly compressible under Elias-Fano's two-level representation. Figure~\ref{fig:compression_alg}(b) shows compression performance on different 100\,M-scale datasets.  ZSTD achieves 51.1\% reduction versus \sysname's 47.0\%, as ZSTD compresses at least 128\,KiB at a time, exploiting cross-vector patterns at a wider context. However, this requires decompressing an entire 128\,KiB block to retrieve a single vector, making it unsuitable for the random per-vector access pattern of ANNS. \sysname compresses each vector independently, preserving fine-grained random access at the cost of a modest compression gap. Compared to Huffman alone, \sysname achieves 13.8\% additional reduction on DecoupleVS100M via XOR-delta encoding; on SIFT100M and SPACEV100M, where 8-bit quantization has already maximized entropy, both methods achieve the same compression ratio, as reported in Exp\#2. 

\subsection{Discussion on Parameter Settings}

\sysname exposes five configurable parameters: (i) chunk size $C$, (ii) segment size, (iii) cache size, (iv) re-ranking batch size $B$, and (v) benefit ratio.  Our default values (\S\ref{subsec:method}) are based on the following principled rationale.  We set $C=4$\,MiB to keep per-chunk metadata overhead $\beta$ (\S\ref{subsec:layout}) within a practical bound across all evaluated datasets (i.e., within 0.1\% of the dataset size). We adopt a segment size of 512\,MiB following conventions in popular vector databases \cite{milvus-segment}. The cache ratio is capped at 1\% of the dataset size, a practical memory footprint that avoids excessive DRAM pressure.  We set $B=10$ to match the result set size $K=10$,  so that each re-ranking batch covers exactly one full result's worth of candidates, balancing I/O cost and re-ranking coverage; we also use the same $B$ for the prefetch stability threshold to align prefetching and re-ranking in latency-aware vector search. The benefit ratio is set to 0.01, reflecting the empirical observation that result-set changes become negligible below this threshold; setting it lower risks unnecessary I/Os while setting it higher risks premature termination. 

\section{Related Work}
\label{sec:related}


\para{Disk-resident ANNS systems.}  ANNS indexes are broadly categorized into graph-based and cluster-based.  Graph-based indexes are pioneered by DiskANN \cite{jayaram19}, which constructs a disk-resident index graph for efficient similarity search; FreshDiskANN \cite{singh21} extends DiskANN to support efficient buffered updates for streaming workloads. Starling \cite{wang24} co-locates neighboring graph vertices on the same disk pages to reduce random I/Os.  PipeANN \cite{guo25} overlaps I/Os and computations during graph traversal via pipelined execution. OdinANN \cite{guo26} supports in-place updates and improves update performance using write-back caching and delta neighbor pruning.  Cluster-based indexes are pioneered by SPANN \cite{chen21}, which partitions vectors into clusters and attaches on-disk posting lists to each centroid for efficient search. SPFresh \cite{xu23} extends SPANN to support efficient in-place streaming updates.  Recent extensions such as SmartANNS \cite{tian24} and FusionANN \cite{tian25} accelerate search by offloading distance computations to SmartSSDs or GPUs.  In contrast, \sysname focuses on improving space efficiency while maintaining high search and update performance. 

Concurrent to our work, DGAI \cite{lou26} and VeloANN \cite{zhao26} both redesign the on-disk storage layout to improve I/O efficiency during search and updates.  DGAI focuses on optimizing update efficiency and query performance through a similarity-aware dynamic layout and hierarchical PQ, without addressing storage efficiency; its decoupled design incurs up to 27.8\% additional storage overhead.  VeloANN introduces a lossy compressed vector layout with affinity-based record co-placement to maximize search throughput; however, its lossy compressed layout does not preserve data fidelity, and it targets static workloads without addressing the impact of per-record updates on search performance.  \sysname targets component-aware lossless compression for persistent vector storage to preserve data fidelity while maintaining high search and update performance. 


\para{Low-storage ANNS.} Disk-resident ANNS systems \cite{jayaram19,tian25} often cache lossy-compressed vectors in memory.  Product quantization (PQ) \cite{jegou10} is the predominant lossy vector compression technique; it represents high-dimensional vectors with compact codes by quantizing sub-dimensions into codebooks. To improve quantization accuracy, Optimized-PQ \cite{ge13} reduces inter-dimensional correlation via orthogonal rotation preprocessing on the original vectors, while RaBitQ \cite{gao24} adopts an integrated design of codebook construction and distance estimation. However, they inevitably incur quantization errors in search.  To meet high recall, existing disk-resident ANNS systems \cite{jayaram19, tian25} still store full-precision vectors on disk for a final re-ranking step.  LEANN \cite{wang26} eliminates full-precision vector storage by recomputing embeddings on the fly during search for low storage overhead.  However, it incurs significant recomputation latency during search.  \sysname applies component-aware lossless compression to persistent vector storage for space savings with negligible decompression overhead.

\para{General-purpose lossless compression.} Lossless compression reduces data volume while ensuring perfect reconstruction.  General-purpose lossless compression methods, such as entropy coders (e.g., Huffman \cite{knuth85}, ANS \cite{duda13}) and dictionary coders (e.g., LZ77 \cite{rigler07} and LZ4 \cite{lz4}) are widely adopted, but overlook data-specific characteristics.  Domain-specific techniques achieve higher compression savings by exploiting inherent data properties, such as AI model weights \cite{hershcovitch25}, LLM storage \cite{wang25}, and sorted inverted lists \cite{ottaviano14, wang17}.  Following this principle, \sysname designs tailored lossless compression for both vector data and auxiliary index metadata, leveraging their distinct data locality patterns to maximize storage savings without degrading performance.

\section{Conclusion}
\label{sec:conclusion}

\sysname addresses the storage compression problem in disk-resident graph ANNS systems through component-aware compression to data-index decoupled storage. With carefully designed lossless compression techniques, data layouts, and search/update paths, \sysname addresses the space, read, and write inefficiencies of existing co-located designs while preserving search and update performance. Evaluation on real-world public and proprietary billion-scale datasets shows that \sysname substantially reduces storage space while delivering improved or competitive search and update performance compared to state-of-the-art disk-resident graph ANNS systems.

\end{sloppypar}

{
\bibliographystyle{plain}
\bibliography{reference}
}

\clearpage
\renewcommand{\thesection}{\Alph{section}} 
\setcounter{section}{0} 
\section{Appendix}
\label{sec:appendix}
\renewcommand{\thesection}{\arabic{section}}
\setcounter{section}{0} 

We report additional evaluation results that supplement the findings in the main paper.

\para{Exp\#9 (P99 tail latency versus accuracy).}  We evaluate the P99 tail latency of \sysname against DiskANN and PipeANN on DecoupleVS100M and SIFT100M, using the same experimental settings as Exp\#4 (\S\ref{subsec:overall}).

\begin{figure}[!h]
\centering
\setlength{\tabcolsep}{0pt}
\begin{tabular}{cc}
\multicolumn{2}{c}{\includegraphics[height=18pt]{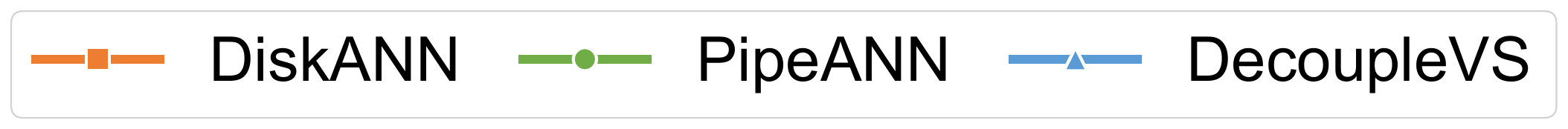}} \\
\multicolumn{2}{c}{\vspace{-16pt}} \\
\includegraphics[width=0.50\linewidth]{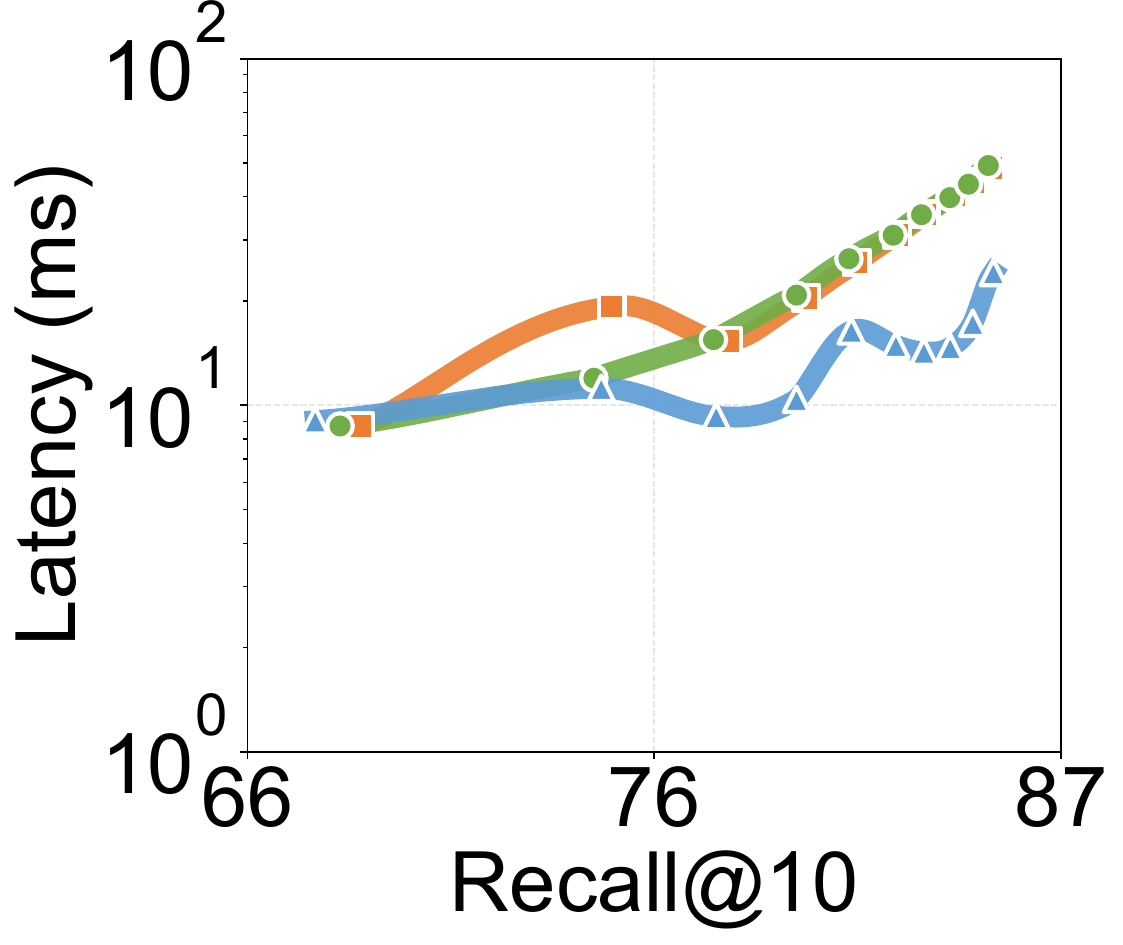} &
\includegraphics[width=0.48\linewidth]{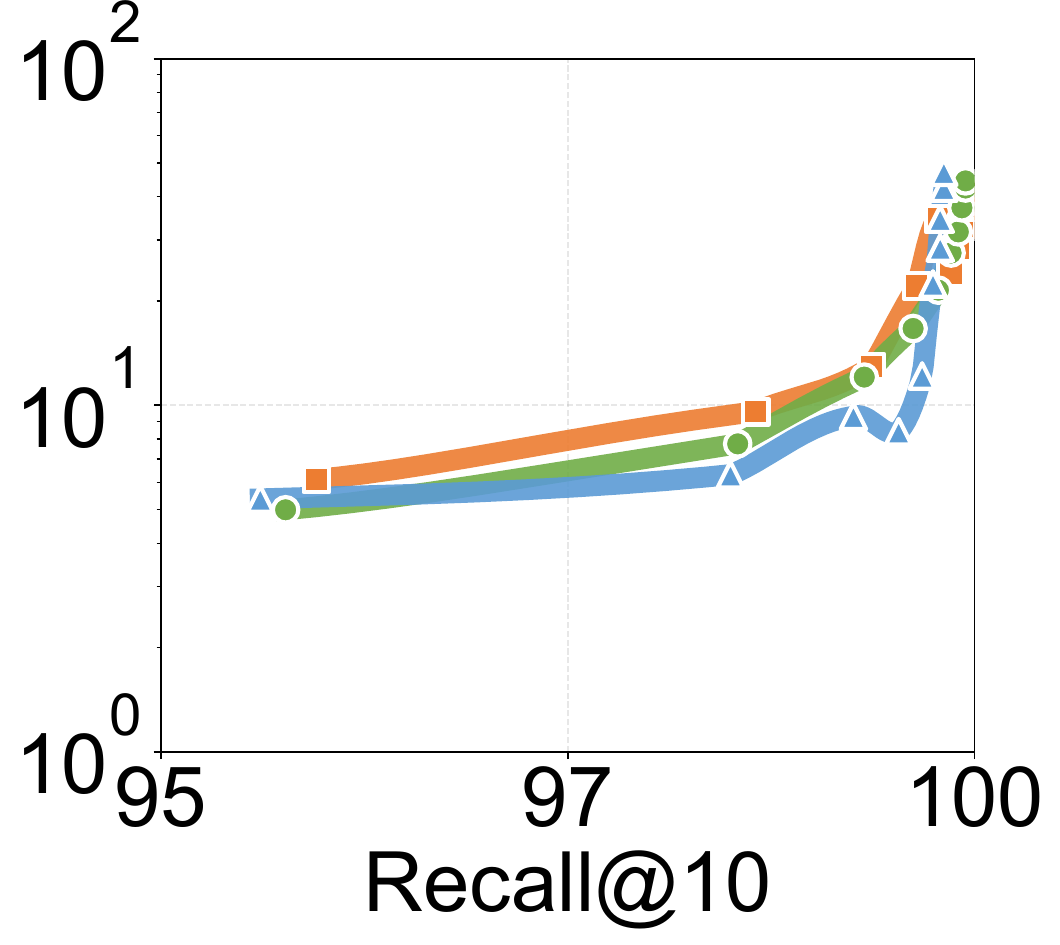}
\vspace{-3pt}\\
\parbox[t]{1.5in}{\centering \small (a) DecoupleVS100M} &
\parbox[t]{1.5in}{\centering \small (b) SIFT100M}
\end{tabular}
\vspace{-9pt}
\caption{Exp\#9 (P99 tail latency versus accuracy). P99 latency (ms) is plotted against Recall@10 (\%); lower is better.}
\label{fig:tail_latency}
\vspace{-6pt}
\end{figure}

Figure~\ref{fig:tail_latency} shows the results.  On DecoupleVS100M, \sysname consistently reduces P99 latency across all recall levels.  At recall@10 $= 83.0\% \pm 0.1\%$, \sysname reduces P99 latency by 51.9\% over both DiskANN and PipeANN.  At recall@10 $= 85.5\% \pm 0.1\%$, the reductions are 50.1\% over DiskANN and 50.9\% over PipeANN.  These improvements are consistent with the average latency reductions reported in Exp\#4, confirming that \sysname's latency-aware search strategy (\S\ref{subsec:search}) reduces not only mean latency but also worst-case tail latency by removing full-precision vector fetching from the critical search path.

On SIFT100M, \sysname reduces P99 latency at moderate recall levels: at recall@10 $= 98.7\% \pm 0.1\%$, \sysname achieves reductions of 34.2\% over DiskANN and 18.5\% over PipeANN.  At very high recall (above 99.8\%), all systems must traverse a larger portion of the graph to meet accuracy targets, reducing the opportunity for adaptive prefetching and narrowing the P99 gap across systems.

\end{document}